%% file: furer.tex
\iffalse
\documentclass{mcom-l}
\else
\documentclass[a4paper]{amsart}
\fi

\input{furer-macros.tex}

\title{Fast integer multiplication using \goodbreak generalized Fermat primes}

\author{Svyatoslav Covanov}
\author{Emmanuel Thomé}
\email{svyatoslav.covanov@loria.fr, emmanuel.thome@inria.fr}
\address{Université de Lorraine, CNRS, INRIA, LORIA, F-54000 Nancy, France}


\date{}

\dedicatory{}

\usepackage{fancyvrb}
\usepackage{tikz}

\def\CodeBox{\FV@Environment{}{CodeBox}}

\def\FVB@CodeBox#1{%
\@bsphack
\begingroup
\FV@UseKeyValues
\gdef\codelanguage{#1}%
\gdef\FV@TheVerbatim{}%
\def\FV@ProcessLine##1{%
\expandafter\gdef\expandafter\FV@TheVerbatim\expandafter{%
\FV@TheVerbatim\FV@ProcessLine{##1}}}%
\gdef\FV@TheVerbatim{}%
\FV@Scan}

\def\FVE@CodeBox{%
\endgroup\@esphack
    \begin{center}
        \begin{tikzpicture}
            \node[rectangle,rounded corners,inner
            sep=6pt,fill=yellow!25!white] (A) {%
                \begin{minipage}{0.9\textwidth}
                    \begin{center}
                        \expandafter\def\expandafter\FV@KeyValues\expandafter{\FV@KeyValues,fontsize=\footnotesize}%
                        \FV@UseVerbatim\FV@TheVerbatim
                    \end{center}
                \end{minipage}%
            };
            \node[fill=white,draw=yellow!25!white,thick,yshift=1ex] at (A.north) {\footnotesize
            \texttt{\codelanguage} code};
        \end{tikzpicture}
    \end{center}
}

\DefineVerbatimEnvironment{CodeBox}{CodeBox}{}
\makeatother

\makeatletter
\def\@technicalL#1{(2#1+2.5\lg{#1})}
\def\@technicalLexp#1{2^{2#1}{#1}^{2.5}}
\def\@oldtechnicalS#1{1+2{#1}^2}
\def\@technicalS#1{{#1}^{2.5}}
\iffalse
\usepackage{highlight}
\def\technicalL#1{\highlight[
                    highlighter colour=red,
                    highlighter cap=round,
                    highlighter opacity=.4
                ]{\@technicalL{#1}}}
\def\technicalLexp#1{\highlight[
                    highlighter colour=red,
                    highlighter cap=round,
                    highlighter opacity=.4
                ]{\@technicalLexp{#1}}}
\def\technicalS#1{\highlight[
                    highlighter colour=blue,
                    highlighter cap=round,
                    highlighter opacity=.4
                ]{\@technicalS{#1}}}
\def\oldtechnicalS#1{\highlight[
                    highlighter width=18pt,
                    highlighter colour=orange,
                    highlighter cap=round,
                    highlighter opacity=.4
                ]{\@oldtechnicalS{#1}}}
\else
\let\technicalS\@technicalS
\let\technicalL\@technicalL
\let\technicalLexp\@technicalLexp
\let\oldtechnicalS\@technicalS
\fi
\makeatother

\begin{document}

\begin{abstract}
    For almost 35 years, Schönhage-Strassen's algorithm has been the
    fastest algorithm known for multiplying integers, with a time
    complexity $O(n \cdot \log n \cdot \log \relax \log n)$ for
    multiplying $n$-bit inputs.  In 2007, Fürer proved that there exists
    $K>1$ and an algorithm performing this operation in $O(n \cdot \log n
    \cdot K^{\log^* n})$. Recent work by Harvey, van der Hoeven, and
    Lecerf showed that this complexity estimate can be improved in order
    to get $K=8$, and conjecturally $K=4$. Using an alternative
    algorithm, which relies on arithmetic modulo generalized Fermat
    primes (of the form $\gfprime r\lambda$), we obtain conjecturally the same result $K=4$ via a careful
    complexity analysis in the deterministic multitape Turing model.
\end{abstract}

\maketitle

\section{Introduction} 
\label{sec:intro}
The first nontrivial algorithm for multiplying $n$-bit integers is
Karatsuba's divide-and-conquer algorithm~\cite{karatsuba-en}, which
reaches the complexity $O(n^{\lg 3})$, with $\lg$ denoting 
the logarithm in base $2$.  The Karatsuba algorithm can be
viewed as a simple case of a more general evaluation-interpolation
paradigm.  In the form of the Toom-Cook algorithm~\cite{1963-toom}, this
paradigm can be extended so as to reach the complexity
$O(n^{1+\epsilon})$ for any $\epsilon>0$.

The first algorithm to achieve what is called \emph{quasi-linear}
complexity is Schönhage and Strassen's~\cite{1971-scho,Schonhage82}. First, the
Schönhage-Strassen algorithm uses the fast Fourier transform (FFT) as a
means to quickly evaluate a polynomial at the powers of a primitive root
of unity~\cite[\S8]{vonzurGathen:1999:MCA:304952}. Second, the complexity is
obtained by an appropriate choice of a ring \cR\ in which this evaluation
is to be carried out. Namely, the choice $\cR=\rquo\bZ{(2^t+1)}$, for
$t$ a suitable power of two, yields the complexity $O(n\cdot \log n \cdot
\log\relax\log n)$, while other natural choices for \cR\ appeared to
yield inferior performance at the time.

In 2007, M. Fürer observed that
the ring $\cR=\rquo{\bC[x]}{(x^P+1)}$, for $P$ a suitable power of two, is
particularly interesting~\cite{DBLP:journals/siamcomp/Furer09}. Using this ring \cR, it is possible to take
advantage of large-radix FFT to obtain the improved complexity
$O(n\cdot\log n\cdot2^{O(\log^*n)})$ (in contrast, radix-2 FFT is
sufficient for the Schönhage-Strassen algorithm). The notation
$\log^*$ denotes the iterated logarithm (see
§\ref{subsec:notations-log}). Fürer's result was an acclaimed
improvement on the complexity of the Schönhage-Strassen algorithm which
had remained unbeaten for 35 years.

Fürer's algorithm, as it stands, is perceived as a theoretical result.
The last decade has seen various articles explore potential improvements
on Fürer's work, either meant to make the complexity more explicit, or
to provide possibly more practical variants. An early extension of
Fürer's work, proposed in~\cite{DKSS2008}, replaces the field \bC\ in the
definition of \cR\ by a $p$-adic ring and reaches an identical asymptotic
complexity. This $p$-adic variant can be expected to ease precision
issues for potential implementations.
Harvey, van der Hoeven and Lecerf in~\cite{HavdHLe16}, and later Harvey and van
der Hoeven in~\cite{HavdH16} propose new algorithms
and a sharper complexity analysis that
allows one to make the complexity more explicit, namely $O(n\cdot \log n
\cdot 8^{\log^* n})$ and even $O(n\cdot \log n\cdot  4^{\log^*
n})$ conjecturally. In comparison, they also show that a careful analysis
of Fürer's original algorithm reaches the complexity $O(n\cdot \log
n\cdot  16^{\log^* n})$

This article presents another variant of Fürer's algorithm. Our algorithm
reaches the complexity $O(n\cdot \log n\cdot  4^{\log^* n})$ and
relies on a conjecture which can be regarded as an explicit version of
the Bateman-Horn conjecture~\cite{batemanhorn1962}, supported by
numerical evidence. Namely, our assumption is as follows.
\begin{recallhypothesis}
    Let $\lambda\geq2$ be an integer.  For any real number $R$ such that
    $2^{\lambda} \leq R \leq 2^{2\lambda}$, there exists a generalized
    Fermat prime $p=\gfprime r\lambda$ such that $R\leq r <
    \technicalS\lambda R$.
\end{recallhypothesis}
The key concept of our algorithm is the use of a
chain of generalized Fermat primes (of the form \gfprime r\lambda) to
handle recursive calls. We therefore differ significantly from the
approach followed by Harvey, van der Hoeven and Lecerf
in~\cite{HavdHLe16,HavdH16}. In a sense
however, some lineage can be drawn between our work and an early article
by Fürer~\cite{ExtAbstrFurer} (from 1989), which is dependent on the
assumption that there exist infinitely many Fermat primes. The latter
assumption, however, is widely believed to be wrong, so our variant fills
a gap here.

The way we obtain a complexity formula with $4^{\log^* n}$ and not
$16^{\log^* n}$ as for Fürer's algorithm is original.  In fact, two
improvements stack onto one another. First, we encode integers
to be multiplied as integers modulo generalized Fermat primes, and not as
polynomials. This saves a factor of two in the sizes of intermediate
products. Second, generalized Fermat primes allow to avoid the Kronecker
substitution, and therefore we use less padding in the intermediate
products.
\smallskip

This article is organized as follows.  Section~\ref{sec:generalities}
reviews classical facts about quasi-linear integer multiplication
algorithms.  Fürer's algorithm in particular is introduced in
Section~\ref{sec:furerbounds}.  Section~\ref{sec:admissible-gfp} studies
generalized Fermat primes, and their relation to
the Bateman-Horn conjecture.
We then proceed to define
a chain of generalized Fermat primes which is crucial to tackle sizes
above a certain threshold.  Section~\ref{sec:algo} uses material
developed in the previous sections and presents our new algorithm (in
fact two algorithms),
with the corresponding recursive complexity equations. We derive an
asymptotic complexity estimate in Section~\ref{sec:solve-rec}. Section~\ref{sec:practical}
discusses how practical our algorithm could be, and proposes projected
timings.
Appendix~\ref{appendix:proofconjc} gives the proof of
Proposition~\ref{prop:lowerbound-C}.
\section{Background}\label{sec:generalities}
\subsection{Notations}
\label{subsec:notations-log}
Throughout the article, $\lg x$ denotes the logarithm in base $2$, and
$\log x$ denotes the natural logarithm. We use the notation $\mlog{m}$
to denote the $m$-th iterate of the $\log$ function, so that
$\mlog{m+1}=\log\circ\mlog m$ (and likewise for $\lg$).

We denote by $\log^*$ the iterated logarithm function, defined recursively
by $\log^*x=0$ for any real number $x\leq1$, and by
$\log^*x=1+\log^*(\log x)$ for $x\geq1$.

The notation $\closedintervalZ{u,v}$ denotes the set of integers $x$ such
that $u\leq x\leq v$.

The notation $u=\Theta(v)$ denotes: $(u=O(v)\text{ and
}v=O(u))$.

\subsection{Integers to polynomials}
\label{subsec:inttopol}

Let $a$ and $b$ be positive $n$-bit integers to be multiplied and
$c=ab$.
Standard substitution techniques (see e.g.\
\cite{Bernstein01m3}) allow one to compute $c$ via the computation of the
product $C(x)=A(x)B(x)$, where $A$ and $B$ are univariate polynomials
related to $a$ and $b$. Polynomials are taken over some well-chosen ring
\cR. Such a procedure is described in
Algorithm~\ref{algo:integer-mul:clump+eval}, where we highlight the
possibility of computing the product $C(x)=A(x)B(x)$ by
multipoint evaluation and interpolation if the ring \cR\ in which
computations take place provides a nice and sufficiently large set of
interpolation points. (In this section, we do not explicitly fix a choice
for \cR. We will do so later on in this article.)

\begin{myalgorithm}{MultiplyIntegersViaMultipointEvaluation}
\caption{%
    \algorithmlabel{algo:integer-mul:clump+eval}\relax
    Multiply in \bZ\ via multipoint evaluation of polynomials}
\tracingonline=0
\tracingmacros=0
\begin{algorithmic}
        \Function{\currentalgorithm}{$a,b,\eta,\cR$}
    \Input{$a$, $b$ two positive $n$-bit integers;\\
        \qquad $\eta$ a power of two; we let $N=\left\lceil
                2n/\lg\eta\right\rceil$\\
        \qquad \cR\ a ring where integers below $N\eta^2$ are
        unambiguously represented\\
        \qquad $\cS\subset\cR$ a set of $N$ evaluation points}
        \Output{$c = a \cdot b$}
        \State Let $A(x)\in\bZ[x]$, with all coefficients in $[0,\eta)$,
        be such that $A(\eta)=a$.
        \State Define $B(x)$
        likewise.
        \State $\hat A \gets
        \Call{MultiEvaluation}{A,\cS}$; define $\hat B$
        likewise.
        \State $\hat C \gets \Call{PointwiseProduct}{\hat
        A,\hat B}$.
        \State $C \gets \Call{Interpolation}{\hat C,\cS}$
        \State
        Reinterpret $C$ as a polynomial in $\bZ[x]$.
        \State \Return $c=C(\eta)$.
        \EndFunction
\end{algorithmic}
\end{myalgorithm}

The procedure followed by Algorithm~\ref{algo:integer-mul:clump+eval} is
in fact quite general, and can be applied to a wider range of bilinear
operations than just integer multiplication. For example, one can imitate
this algorithm to multiply polynomials or power series in various rings,
or to compute other operations such as middle products or dot products.
The latter example of the dot product is archetypal of the situation
where results of the \textproc{MultiEvaluation} step (as e.g.\ $\hat A$ in
Algorithm~\ref{algo:integer-mul:clump+eval}) are used more than once. The
conditions on \cR\ that are used to guard against possible overflow must
be adjusted accordingly.

\subsection{Cooley-Tukey FFT}
\label{subsec:cooley-tukey}

We now discuss how multi-evaluation can be performed efficiently. This
depends first and foremost on the number of evaluation points $N$ and on
the ring \cR. FFT algorithms are special-purpose algorithms adapted to
evaluation points chosen among roots of unity in \cR. In order to allow
\cR\ to be a non-integral ring, we need the following definition.

\begin{definition}
\label{def:principalroot}
Let $N\geq1$ be an integer, and \cR\ be a ring of characteristic zero or
    characteristic coprime
    to $N$, containing an $N$-th root of unity $\omega$.
We say that $\omega$ is a principal $N$-th root of unity if $\forall i
    \in \closedintervalZ{1,N-1},
            \sum_{j=0}^{N-1}\omega^{ij} = 0.$
\end{definition}
The notion of principal root of unity is stricter than the classical
notion of primitive root, and provides the suitable generalization to
non-integral rings. For example in $\bC\times\bC$, the element
$(1,i)$ is a primitive $4$-th root of unity but not a principal $4$-th
root of unity.

Using the set of powers of $\omega$ as a set of evaluation points, we
define the discrete Fourier transform (DFT).

\begin{definition}[Discrete Fourier Transform (DFT)]
Let \cR\ be a ring with $\omega$ a principal $N$-th  root of unity.
The DFT of length $N$ and base root $\omega$ over \cR\ is the
    ring isomorphism $\DFT_{N,\omega}$ defined as:
    $$\left\{
        \begin{array}{rcl}
            \quotient{\cR[x]}{(x^N-1)}
            &\rightarrow
            &\quotient{\cR[x]}{(x-1)}
            \times\quotient{\cR[x]}{(x-\omega)}
            \times\cdots
            \times\quotient{\cR[x]}{(x-\omega^{N-1})}\\
            P
            &\mapsto
            &(P(1),P(\omega),\ldots,P(\omega^{N-1})).
        \end{array}\right.$$
\end{definition}

We customarily write a DFT of length $N$ of a polynomial $P$ as the
polynomial $\hat P$ of degree at most $N-1$ defined as $$\hat
P=\DFT_{N,\omega}(P)=P(1)+XP(\omega)+\cdots+X^{N-1}P(\omega^{N-1}).$$
Cooley and Tukey showed in~\cite{Cooley1965} how a DFT of composite order
$N=N_1N_2$ can be computed. This algorithm is also sometimes called
``matrix Fourier algorithm'', alluding to the fact that it performs $N_2$
``column-wise'' transforms of length $N_1$, followed by $N_1$
``row-wise'' transforms of length $N_2$. It is described in
Algorithm~\ref{algo:matrix-fourier}. We note that
Algorithm~\ref{algo:matrix-fourier} implicitly rearranges data (e.g.\ when
computing $B_j$ and $S_i$), and some
work is needed to perform the required matrix transpositions in a
satisfactory way on a multitape Turing machine. Using an algorithm proposed in~\cite{bostan:inria-00103401}, it is
shown in~\cite[§2]{HavdHLe16} that this extra cost is small enough that
it is subsumed within the cost of multiplications by roots of unity in
\cR.

\begin{myalgorithm}{CooleyTukeyFFT}
    \caption{%
    \algorithmlabel{algo:matrix-fourier}
        General Cooley-Tukey FFT of order $N=N_1N_2$}
    \begin{algorithmic}
    \Function{\currentalgorithm}{$N_1$,$N_2$,$\omega$,$A$}
        \Input{$A = \sum_{i=0}^{N-1}a_i X^i\in\rquo{\cR[X]}{(X^N-1)}$;\\[]
            \qquad $\omega$ a principal $N$-th root of unity}.
            Let $\omega_1=\omega^{N_2}$ and $\omega_2=\omega^{N_1}$.
    \Output{$\hat A = \DFT_{N,\omega}(A) = A(1)+A(\omega)X+\cdots+A(\omega^{N-1})X^{N-1}$}
        \State Let $(B_j(X))_j\in\cR[X]^{N_2}$ be such that
        $A(X)=\sum_{j<N_2}B_j(X^{N_2})X^j$
    \For{$j \in \closedintervalZ{0,N_2-1}$}
        \State $\hat{B_j} \gets \DFT_{N_1,\omega_1}(B_j)$
        \Comment{$\omega_1=\omega^{N_2}$ is a principal $N_1$-st root}
        \State $
        \Gamma
        _j \leftarrow \hat{B_j}(\omega^j X)$
        \Comment{$(\Gamma_j)_j$ are ``twisted transforms'' of the
        $(B_j)_j$}
\EndFor
        \State Let $(S_i(Y))_i\in\cR[Y]^{N_1}$ be such that
        $\sum_{j < N_2}\Gamma
        _j(X) Y^j=\sum_{i < N_1} S_i(Y) X^i$
\For{$i \in \closedintervalZ{0,N_1-1}$}
        \State $\hat{S_i} \gets \DFT_{N_2,\omega_2}(S_i)$
        \Comment{$\omega_2=\omega^{N_1}$ is a principal $N_2$-nd root}
\EndFor
        \State \Return $\sum_{i<N_1} \hat{S_i}(X^{N_1})X^i$
\EndFunction
\end{algorithmic}
\end{myalgorithm}

The notation $\DFT_{N,\omega}$ denotes a mathematical object rather than
an algorithm. Therefore, we need to detail how recursive computations of
$\DFT_{N_1,\omega_1}$
and $\DFT_{N_2,\omega_2}$
are handled in Algorithm~\ref{algo:matrix-fourier}.
Two approaches are rather typical instantiations of the Cooley-Tukey
algorithm
when the length $N$ is a power of two:
\begin{itemize}
    \item ``radix-two FFT'': For a length $N=2^k$, compute $N_2=2^{k-1}$
        transforms of length $N_1=2$ (often called ``butterflies''), then
        recurse with two transforms of length $2^{k-1}$. We use the
        notation $\textproc{Radix2FFT}(N,\omega,A)$ for this
        algorithm.
    \item ``large-radix FFT'': More generally, for a length $N=2^{uq+r}$
        with $r<u$, and $q>0$, compute $N_2=N/2^u$ transforms of length
        $N_1=2^u$, then
        recurse with transforms of length $N_2=N/2^u$. When all recursive
        calls are unrolled, we see that the computation 
        is based on transforms of length $N_1=2^u$ (or
        $N=2^r$ at the very end of the recursion). Those are done with
        \textproc{Radix2FFT}. We use the notation
        $\textproc{LargeRadixFFT}(N,\omega,2^u,A)$ for this
        algorithm. 
\end{itemize}
It is clear that the latter approach specializes to the former when
$u=1$.

Large-radix FFT is often used for practical purposes, as it typically
improves application performance.  As we observe later on in this
article, this has a stronger impact in the context of Fürer's algorithm,
since the overall complexity is very dependent on this technique.

The computational interest of using FFT algorithms for multi-evaluation
follows from the count $C(N)$ of operations in \cR\ that are required for an FFT of
length $N=2^k$. Using radix $2^u$ as an example ($u$ being a constant),
we have $C(N)/N=C(2^u)/2^u+C(N/2^u)/(N/2^u)+O(1)$, from which it follows
that asymptotically we have $C(N)=O(N\lg N)$.

\smallskip
Two additional comments are worth mentioning. First, we define a similar
isomorphism, denoted
$\HDFT_{N,\omega}$, by the multi-evaluation at odd powers of a $2N$-th root
of unity $\omega$:
    $$\left\{
        \begin{array}{rcl}
            \quotient{\cR[x]}{(x^N+1)}
            &\rightarrow
            &\quotient{\cR[x]}{(x-\omega)}
            \times\quotient{\cR[x]}{(x-\omega^3)}
            \times\cdots
            \times\quotient{\cR[x]}{(x-\omega^{2N-1})}\\
            P
            &\mapsto
            &(P(\omega),P(\omega^3),\ldots,P(\omega^{2N-1})).
        \end{array}\right.$$
A half-DFT of length $N$ can be computed at
the same cost as a DFT of length $N$, plus $N$ extra multiplications for
scaling.
More precisely, to multi-evaluate $P(X)$ (an element of
$\quotient{\cR[x]}{(x^N+1)}$) at
$\omega,\omega^3,\ldots,\omega^{2N-1}$, we compute
$\HDFT_{N,\omega}(P(X))=\DFT_{N,\omega^2}(P(\omega X))$.
Half-DFTs are used for polynomial products modulo $X^N+1$, as
opposed to $X^N-1$. Such convolutions are called \emph{negacyclic}.

Also, it is straightforward to verify that the task of interpolating a
polynomial $A$ from its multi-evaluation $\hat A$ can be done with
essentially the same algorithm (see e.g.\
\cite[§8]{vonzurGathen:1999:MCA:304952}). The inverse transforms are
written as \begin{align*}
    \operatorname{IFT}_{N,\omega}(A(X)) &=
    \frac1N\DFT_{N,\omega^{-1}}(A(X)),\\
    \operatorname{Half-IFT}_{N,\omega}(A(X)) &=
    \frac1N\operatorname{IFT}_{N,\omega^2}(A(X))(\omega^{-1}X).
\end{align*}
We shall not discuss this point further.
\subsection{Complexity of integer multiplication}
\label{subsec:cplx-intmul}

\begin{notation}
We denote by $\sM(n)$ the cost of the multiplications of two $n$-bit
integers in the deterministic multitape Turing model~\cite{1994-papadimitriou},
also called bit complexity.
\end{notation}

By combining the evaluation-interpolation scheme of
§\ref{subsec:inttopol} with FFT-based multi-evaluation and
interpolation as in §\ref{subsec:cooley-tukey},
we obtain quasi-linear integer multiplication algorithms.  We identify
several tasks whose cost contributes to the bit complexity of such algorithms.
\begin{itemize}
    \item conversion of the input integers to polynomials
        in $\cR[X]$;
    \item multiplications by roots of unity in the
        FFT computation;
    \item linear operations in the FFT computation (additions, etc);
    \item pointwise products of elements of
        \cR.
    \item recovery of the resulting integer from the computed polynomial.
\end{itemize}

Algorithm~\ref{algo:integer-mul:clump+eval} chooses $\eta$ a power of two
so that the first and last steps above have linear complexity (at least
provided that elements in \cR\ are represented in a straightforward way).
If we go into more detail, $\sM(n)$ then expresses as $\sM(n) = C(N)\cdot
K_{\text{FFT}}(\cR) + N\cdot K_{\text{PW}}(\cR) + O(n)$, with the following
notations.
\begin{itemize}
    \item $K_{\text{FFT}}(\cR)$ denotes the cost for the multiplication
        by powers of $\omega$ in \cR\ that occur within the FFT
        computation.
    \item $K_{\text{PW}}(\cR)$ denotes the binary cost for the pointwise
        products in \cR.
\end{itemize}
The costs $K_{\text{PW}}(\cR)$ and $K_{\text{FFT}}(\cR)$ are not
necessarily equal.
        Of course, both may involve recursive calls to fast multiplication
        algorithms.
\subsection{Choice of the base ring}
\label{subsec:choice-ring}

Depending on \cR, the bit complexity estimates of
§\ref{subsec:cplx-intmul} can be made more precise.  Some rings have
special roots of unity that allow faster operations (multiplication,
in \cR, most importantly) than others.  Several choices for \cR\ are
discussed in~\cite{1971-scho}. We describe their important
characteristics when the goal is to multiply two $n$-bit integers.

The choice $\cR=\bC$ might seem natural because roots of unity are
plenty. The precision required calls for some analysis.
\begin{itemize}
    \item A precision of $t=\Theta(\lg n)$ bits
        is compatible with a
        transform length $N=\Theta({n}/{\lg n})$
        (see~\cite[§3]{1971-scho}), in the sense that the polynomials that
        we multiply can be represented on $tN$ bits and the product would
        not be correct if $t$ were smaller (thus, $t= \Theta(\lg n)$ is
        optimal).
    \item Costs for operations in \cR\ are $K_{\text{FFT}}(\cR)=
        K_{\text{PW}}(\cR)= O(\sM(\lg n))$.
\end{itemize}
This yields $\sM(n)=O(N\lg N\cdot  \sM(\lg n))=O(n \cdot \sM(\lg n))$, so
that
$$\sM(n)=2^{O(\lg^*  n)}\cdot n\cdot\lg n\cdot\lg\lg
n\cdot\lg\lg\lg n\cdot\ldots,$$ where the number of recursive calls is
$\lg^*  n+O(1)$.
\smallskip

Schönhage and Strassen (originally in~\cite{1971-scho}, later changed to a
simpler variant
in~\cite{Schonhage82}) proposed the alternative $\cR=\ZnZ{(2^t+1)}$, in
which $2$ is a principal $2t$-th root of unity.
Their algorithm
multiplies $n$-bit integers modulo $2^n+1$, for suitable $n$ (to fix
ideas, take $n$ a power of two).
This algorithm can be adapted to the general integer multiplication by
multiplying $n$-bit integers modulo $2^{2n}+1$.
\begin{itemize}
    \item 
        We pick a transform length $N$ slightly below $\sqrt n$, and divide both inputs in chunks of
        $\lceil n/N\rceil$ bits.
    \item We choose the ring $\cR=\ZnZ{(2^t+1)}$ with $t$ subject to
        several constraints, namely that $t$ be a multiple of $N$, and
        that $t\geq 2n/N + \log_2 N + O(1)$.
        The algorithm uses a negacyclic convolution in
        $\rquo{\cR[X]}{(X^N+1)}$.
    \item The cost $K_{\text{FFT}}(\cR)$ is linear in $t$, as all
        multiplications by power of $\omega$ reduce to binary shifts. We
        thus have $K_{\text{FFT}}(\cR) = O(\sqrt n)$.
    \item The cost $K_{\text{PW}}(\cR)$ is the cost of a recursive
        multiplication
        modulo $2^t+1$.
	Thus, $K_{\text{PW}}(\cR)=\sM(t)$.
\end{itemize}
For the complexity analysis, write
$m(n)=\sM(n)/(n\lg n)$. We then have $m(n)=O(1)+(1+o(1))\cdot\frac{2\lg t}{\lg
n}m\left(t\right)$. Dealing with $(1+o(1))$ 
with due care (see in particular~\cite{Schonhage82}), we eventually obtain
$\sM(n)=O(n\cdot\log
n\cdot\log\relax\log n)$.
\smallskip

\section{Fürer-type bounds}\label{sec:furerbounds} 
The choices mentioned in §\ref{subsec:choice-ring} have orthogonal
advantages and drawbacks. The complex field allows larger transform
length, shorter recursion size, but suffers, when looking at the cost
$K_{\text{FFT}}(\bC)$, from expensive roots of unity.  Those account for
the term $\log_2 n \cdot \mlg2 n \cdots \mlg{\log^* n} n$ in the
complexity of the multiplication of $n$-bit integers using this base
ring.

Fürer proposed two distinct algorithms: one in~\cite{ExtAbstrFurer} and, some
20 years later, in~\cite{DBLP:journals/siamcomp/Furer09}.  The scheme
proposed in~\cite{ExtAbstrFurer} relies on the assumption that there
exist
infinitely many Fermat primes, which is unfortunately widely believed to
be wrong.
We briefly review here the algorithm proposed later
in~\cite{DBLP:journals/siamcomp/Furer09}.


\subsection{A ring with convenient roots of unity}
\label{subsec:polring}
Fürer proposed in~\cite{DBLP:journals/siamcomp/Furer09} to use the ring
$\cR=\quotient{\bC[x]}{(x^{2^{\lambda}}+1)}$, which has a natural principal 
${2^{\lambda+1}}$-th root of unity, namely $x$. Notice that
\cR\ is also isomorphic to $\prod_{j=0}^{2^\lambda-1}\cR_j$, where the
component $\cR_j$ is \quotient{\bC[x]}{(x-\exp((2j+1)i\pi/2^\lambda))}.
For any integer $N$ which is a multiple of $2^{\lambda+1}$ (and in
particular for powers of two of higher order), we define $\omega_N$ as the
unique element of \cR\ that maps to $\exp(2(2j+1)i\pi/N)$ in $\cR_j$.
Lagrange interpolation can be used to compute $\omega_N$ explicitly. We verify
easily that:
\begin{itemize}
    \item $\omega_N$ is a principal $N$-th root of unity.
    \item $\omega_N^{N/2^{\lambda+1}}$ maps to $x=\exp((2j+1)i\pi/2^\lambda)$
        in $\cR_j$, so that $\omega_N^{N/2^{\lambda+1}}=x$ in \cR.
\end{itemize}
The latter point implies that among powers of $\omega_N$, some enjoy
particularly easy operations.

Consider now how an FFT of length $N=2^{(\lambda+1)\cdot q+r}$ can be
computed with \algorithmref{algo:matrix-fourier}. For $q>0$, we write
$N=N_1N_2$ with $N_1=2^{\lambda+1}$. This way,
Algorithm~\ref{algo:matrix-fourier} calls an external algorithm (say,
radix-two FFT) for the transform of length $N_1=2^{\lambda+1}$, and calls
itself recursively for the transform of length $N_2$. The key observation
is that in the many transforms of length $N_1$ that are computed within
the recursion, multiplications by roots of unity are then multiplications
by powers of $x\in\cR$, and therefore inexpensive. We can count the 
remaining multiplications that occur within the
recursion. We call them ``expensive'' although in truth some might
actually be accidentally cheap. Those correspond to the scaling operation $\Gamma_j
\leftarrow \hat{B_j}(\omega^j X)$ in Algorithm~\ref{algo:matrix-fourier}.
Their count $E(N)$ satisfies $E(N)=2^{\lambda+1}
E({N}/{2^{\lambda+1}}) + N$,
from which it follows that $E(N) = N(\lceil\log_{{2^{\lambda+1}}}N\rceil -1)$.
\subsection{Impact on the complexity of integer multiplication}
To multiply integers of at most $n$ bits, where $n$ is a power of two, Fürer selects
${2^{\lambda}}=2^{\lceil\lg\lg n\rceil}$ and proves that precision
$O(\log n)$ is sufficient for the coefficients of the elements of \cR\
that occur in the computation.  The integers to be multiplied are split
into pieces of $2^{2\lambda-1}$ bits. Each piece of $2^{2\lambda-1}$ bits
is transformed into a polynomial of degree $2^{\lambda-1}$ whose
coefficients are encoded on $2^{\lambda}$ bits.  These polynomials are
seen as elements of \cR.  Moreover, the transform length is $N \leq
4n/\lg^2 n$. This decomposition is described in
\algorithmref{algo:multfurer}\footnote{In
line~\ref{algoline:FCM:liftC} of Algorithm~\ref{algo:multfurer}, the
rounding is an acknowledgement that complex numbers may be
represented with restricted precision: if we were to reason only on the
mathematical definition of $\hat A$, $\hat B$, $\hat C$, and $C$, we
could be content with the observation that $C$ has integer
coefficients.}.
\begin{myalgorithm}{FurerComplexMul}
    \caption{%
        \algorithmlabel{algo:multfurer}
        Multiplication of integers with Fürer's algorithm}
    \begin{algorithmic}[1]
	\Function{\currentalgorithm}{$a$,$b$,$n$}
	\Input{$a$ and $b$ two positive $n$-bit integers, where
        $n$ is a power of two}
	\Output{$a\cdot b \mod 2^{2n}+1$}
		\State Let $\lambda = \lceil \mlg2 n\rceil$,
                $\eta = 2^{2^{2\lambda-1}}$,
		$N = 2n/\lg\eta=n/{2^{2\lambda-2}}$,
		\State Let $\cR =
                \quotient{\bC[x]}{(x^{2^\lambda}+1)}$, and
                $\omega=\omega_{2N}$ as in~§\ref{subsec:polring}.
                \State Let $A_0(X)\in\bZ[X]$, with all coefficients in
                $[0,\eta)$, be such that $A_0(\eta)=a$.
                \State Let $\tilde A(X,x)\in\bC[X,x]$, with all
                coefficients integers in $[0,2^{2^\lambda})$, be 
                such that $\tilde A(X,2^{2^\lambda}) = A_0(X)$.
                \Comment $\deg_X\tilde A<N/2$,\ \ $\deg_x\tilde
                A<2^{\lambda-1}$.
                \State Define $B_0(X)$ and $\tilde B$ likewise.
                \State Map $\tilde A$ and $\tilde B$ to polynomials $A$
                and $B$ in
                \quotient{\cR[X]}{(X^N+1)}
                \State $\hat A \gets
                \textproc{LargeRadixFFT}(N,\omega^2,2^{\lambda+1},A(\omega
                X))$ \Comment $\hat A=\HDFT_{N,\omega}(A)$
                \State $\hat B \gets
                \textproc{LargeRadixFFT}(N,\omega^2,2^{\lambda+1},B(\omega
                X))$ \Comment $\hat B=\HDFT_{N,\omega}(B)$
		\State $\hat C \gets
                \textproc{PointwiseProduct}(\hat A,\hat B)$
		\State $C \gets\frac1N
                \textproc{LargeRadixFFT}(N,\omega^{-2},2^{\lambda+1},\hat
                C)(\omega^{-1}X)$
                 \Comment 
                 $C=\operatorname{Half-IFT}_{N,\omega}(\hat C)$
                \State
                \label{algoline:FCM:liftC}
                Lift $C$ to $\tilde
                C\in\bC[X,x]$ with $\deg_X\tilde C<N$, $\deg_x\tilde
                C<2^\lambda$, and integer coefficients (rounding if necessary).
                \State \Return $\tilde{C}(\eta,2^{2^\lambda})$
	\EndFunction
\end{algorithmic}
\end{myalgorithm}

Some non-trivial multiplications by elements of \cR\ are needed in
Algorithm~\ref{algo:multfurer}: 
$3E(N)$ multiplications in recursive calls,
and $4N$ multiplications by scaling factors (because of the negacyclic
convolution) and pointwise products. For
these, we use Kronecker substitution:
we encode elements of \cR\ as integers of bit length $O((\lg n)^2)$, and
then call recursively \textproc{FurerComplexMul}. Other
multiplications by roots of unity 
are cheap.
Their number is
$O(N\log N)$, and their cost is linear in the size of elements of \cR,
that is $O(2^\lambda\log n)$. Additionally, all implicit rearrangement
costs of Algorithm~\ref{algo:multfurer} (see §\ref{subsec:cooley-tukey})
are also within this same bound.
We get the following equation for $\sM(n)$:
\begin{equation}
    \label{eq:cplx-furer}
\sM(n) = N(3\lceil \log_{{2^{\lambda+1}}} N \rceil +1 )\cdot \sM(O(\log
    n)^2) + O(N\log N \cdot {2^{\lambda}}\log n).
\end{equation}

Fürer proves that this recurrence leads to $\sM(n) \leq n\log n
(2^{d\log^*  \sqrt[4]{n}}-d')$ for some constants $d,d' >0$, so that $$\sM(n) =
n\cdot \log n \cdot 2^{O(\log^* n)}.$$

Various directions improve on the above complexity. One of
them is to take advantage of precomputations of transforms of roots of
unity. Briefly put, this transforms the constant $3$
in Equation~\eqref{eq:cplx-furer} to $2$. We do not detail here how this can
be done. In fact, this precomputation strategy is one of the ingredients
(but certainly not the most original one) that the present article
develops in §\ref{sec:algo} to obtain an improved complexity.

As mentioned in §\ref{sec:intro},
Harvey, van der Hoeven and Lecerf in~\cite{HavdHLe16}, and Harvey and van
der Hoeven in~\cite{HavdH16} propose other ways to obtain a better
complexity. They propose new algorithms that achieve complexity
bounds similar to the one that Fürer gets, and improve on the constant~$d$.
Their improvement yields an asymptotic equation similar to the improvement in this
article.  The algorithms in~\cite{HavdHLe16,HavdH16} rely on Bluestein's
chirp transform~\cite{Bluestein70}. They are unrelated to the present
work, and will not be detailed.

\section{Admissible generalized Fermat numbers and primes}\label{sec:admissible-gfp}
 
This section defines admissible generalized Fermat numbers. Our main use
case will be when such
numbers are prime, and we define
a descending chain of such primes.
This section is independent of the previous sections.

\begin{definition}
    \label{not:gfn}
    A \emph{generalized Fermat
    number}
    is an integer of the form
    $r^{2^\lambda}+1$, where
    $\lambda$ and $r$ are two positive integers. We use 
    the shorthand notation $P(r,\lambda)$ for such numbers.
\end{definition}

For notational ease, throughout this
article, whenever we mention a generalized Fermat number $p$, we actually
consider the pair $(r,\lambda)$ rather than the number $p$ alone.  For
this reason, it shall be understood without further mention that $r$ and
$\lambda$ are implicit data that is unequivocally attached to $p$, which
is underlined by the fact that we favor the expression ``let
$p=P(r,\lambda)$ be a generalized Fermat number''.

\subsection{Abundance of generalized Fermat primes}

Asymptotically, the existence of generalized Fermat \emph{primes}
in integer intervals can be obtained via the Bateman-Horn
conjecture~\cite{batemanhorn1962}.
For real numbers $A<B$ and an integer $\lambda\geq1$, we let
$\Delta(\lambda,A,B)$ denote the number of integers
$r\in\closedopeninterval{A,B}$
such that
    $p=f(r)=P(r,\lambda)=r^{2^\lambda}+1$ is a generalized Fermat prime.
The following lemma captures the
asymptotic behaviour of $\Delta$ in specific intervals.  However it will be of little use \emph{per se} but
to define some notations.

\begin{lemma}
    \label{lemma:exists-gfp}
    Fix an integer $\lambda\geq1$. Let $\alpha>1$ be a real number (possibly
    depending on $\lambda$). If the Bateman-Horn conjecture holds for
    $f(x)=x^{2^{\lambda}}+1$,
    then
    $$
        \Delta(\lambda,R,\alpha R) \sim
        \frac{C_\lambda}{2^\lambda}(\li(\alpha R)-\li(R))
        $$
        {as $R\rightarrow\infty$, where we used the notations:}
        \begin{eqnarray*}
        \li(x)=
        \int_2^{x}\frac{\mathrm d
        t}{\log t},
            &
            \displaystyle
        C_\lambda=\frac12\prod_{{p\
    \text{prime}}}\frac{1-{\chi_\lambda(p)}/p}{1-1/p},
            &
        \chi_\lambda(p)=\left\{
        \begin{array}{ll}
            2^\lambda & \text{if $2^{\lambda+1}\mid p-1$},\\
            0 & \text{otherwise.}
        \end{array}
        \right.
        \end{eqnarray*}
\end{lemma}
\begin{proof}
    Bateman and Horn
    \cite{batemanhorn1962} define the constant $C_\lambda$ as above, and
    conjecture
    that 
    as $R$ grows, we have
    $$\Delta(\lambda,1,R)\sim \frac{C_\lambda}{2^\lambda}\li(R)
    \sim \frac{C_\lambda}{2^\lambda}\cdot \frac{R}{\log R}.$$
    Let $\epsilon>0$. Assuming the Bateman-Horn conjecture holds, we have
    for $R$ large enough
    \begin{align*}
        \left\lvert\frac{\Delta(\lambda,R,\alpha
        R)}{\frac{C_\lambda}{2^\lambda}\left(\li(\alpha
        R)-\li(R)\right)}-1\right\rvert<\epsilon\cdot \frac{1+\li(R)/\li(\alpha
        R)}{1-\li(R)/\li(\alpha R)}.
    \end{align*}
    Now since $\li(x)\sim x/\log x$ and $\alpha>1$, the right-hand side
    above converges to a positive constant as $R\rightarrow\infty$. This proves the claim.
\end{proof}

We now go through several steps to provide heuristic arguments supporting
the existence of sufficiently many generalized Fermat primes in our
ranges of interest. Our attention first goes to the asymptotic estimate
on the right-hand side in Lemma~\ref{lemma:exists-gfp}, and to how it
evolves as $\lambda\rightarrow\infty$, for some specific choices of $\alpha$ and
$R$. Table~\ref{tab:C-constant} indicates some experimental values for
the constant $C_\lambda$ (the same data has also been collected
by~\cite{Dubner01distributionof}). While the observation of
Table~\ref{tab:C-constant} would support the empirical claim that
$C_\lambda$ increases as $\lambda$ increases, a proof of such a statement
has eluded us. In Appendix~\ref{appendix:proofconjc}, we prove
Proposition~\ref{prop:lowerbound-C} below, which is a much
weaker statement. We then choose  $\alpha$ and
$R$ so
that the estimate of Lemma~\ref{lemma:exists-gfp} can be shown to tend to
infinity (Proposition~\ref{prop:estimate-is-big}).

\begin{restatable}{proposition}{conjC}
    \label{prop:lowerbound-C}
    Let $C_\lambda$ be as in Lemma~\ref{lemma:exists-gfp}. We have
    $\frac1\lambda=O(C_\lambda)$.
\end{restatable}
\begin{proof}
See Appendix~\ref{appendix:proofconjc}
\end{proof}

\begin{table}
    \begin{center}
        \begin{minipage}{.15\textwidth}

            \begin{tabular}{c|c}
                $\lambda$ & $C_\lambda$\\\hline
                1 & 1.37\\
                2 & 2.68\\
                3 & 2.09\\
                4 & 3.67\\
            \end{tabular}
        \end{minipage}\hfil
        \begin{minipage}{.15\textwidth}
            \begin{tabular}{c|c}
                $\lambda$ & $C_\lambda$\\\hline
                5 & 3.61\\
                6 & 3.94\\
                7 & 3.11\\
                8 & 7.43\\
            \end{tabular}
        \end{minipage}\hfil
        \begin{minipage}{.15\textwidth}
            \begin{tabular}{c|c}
                $\lambda$ & $C_\lambda$\\\hline
                9 & 7.49\\
                10 & 8.02\\
                11 & 7.23\\
                12 & 8.43\\
            \end{tabular}
        \end{minipage}\hfil
        \begin{minipage}{.15\textwidth}
            \begin{tabular}{c|c}
                $\lambda$ & $C_\lambda$\\\hline
                13 & 8.47\\
                14 & 8.01\\
                15 & 5.80\\
                16 & 11.20\\
            \end{tabular}
        \end{minipage}\hfil
        \begin{minipage}{.15\textwidth}
            \begin{tabular}{c|c}
                $\lambda$ & $C_\lambda$\\\hline
                17 & 11.00\\
                18 & 13.01\\
                19 & 13.06\\
                20 & 14.45\\
            \end{tabular}
        \end{minipage}
        \smallskip

            \caption{\label{tab:C-constant}Approximations
            of the infinite product $C_\lambda$, as defined by
            Lemma~\ref{lemma:exists-gfp}. The computation was done by
            enumerating all primes below $10^{11}$, with resulting values
            rounded to nearest. Proposition~\ref{prop:lowerbound-C} shows
            that $\frac{1}{\lambda} = O(C_\lambda)$.
            }
    \end{center}
\end{table}

\begin{proposition}
    \label{prop:estimate-is-big}
    We use the same notations as in Lemma~\ref{lemma:exists-gfp}.
    Let $a(\lambda)$ be a real-valued function such that
    $a(\lambda)\geq\kappa\lambda^{2+\epsilon}$
    for two positive constants
    $\kappa,\epsilon$.
    Then the asymptotic estimate of
    Lemma~\ref{lemma:exists-gfp}, when formulated for $R=2^\lambda$ and
    $\alpha=a(\lambda)$
    is such that, as $\lambda\rightarrow\infty$:
    $$\frac{C_\lambda}{2^\lambda}\left(\li(a(\lambda)\cdot
    2^\lambda)-\li(2^\lambda)\right)\longrightarrow\infty.$$
\end{proposition}
\begin{proof}
    A lower bound for $\left(\li(\alpha
        R)-\li(R)\right)$ is $(\alpha-1)R/\log(\alpha R)$. We have
        $$
            \frac{C_\lambda}{2^\lambda}\left(\li(a(\lambda)\cdot
            2^\lambda)-\li(2^\lambda)\right)
            \geq \frac{C_\lambda}{2\log
            2}\cdot\frac{\kappa\lambda^{2+\epsilon}-1}{\lambda}
            \geq {\lambda C_\lambda}\cdot\frac
            {\kappa \lambda^\epsilon-1/\lambda^2}{2\log 2}.
        $$
        The claim follows, since
        $\frac1\lambda=O(C_\lambda)$ implies that $\lambda C_\lambda
        \lambda^\epsilon$
        tends to $\infty$.
\end{proof}

Our heuristic claim is that for $\alpha=\technicalS\lambda$ (which
fulfills the conditions of Proposition~\ref{prop:estimate-is-big}), the estimate of
Lemma~\ref{lemma:exists-gfp} is accurate enough, as early as for
$R=2^\lambda$.
\begin{hypothesis}
\label{conj:second}
    Let $\lambda\geq2$ be an integer.  For any real number $R$ such that
    $2^{\lambda} \leq R \leq 2^{2\lambda}$, we have $\Delta(\lambda, R,
    \technicalS\lambda R)\geq1$. In other words, there exists a generalized
    Fermat prime $p=P(r,\lambda)$ such that $R\leq r < \technicalS\lambda
    R$.
\end{hypothesis}

Both the constant $C_\lambda$, as well as the accordance of the prime
count $\Delta(\lambda, 1, B)$ with the asymptotic estimate given by the
Bateman-Horn conjecture, have been studied
by~\cite{Dubner01distributionof}. While the experiments
of~\cite{Dubner01distributionof} do support the validity of the
Bateman-Horn conjecture even for primes not very large, we provide
independent experimental data to support Hypothesis~\ref{conj:second}.
We computed
numerically the value $\Delta(\lambda, 2^\lambda, \technicalS\lambda
2^\lambda)$,
as well as the estimate given by
Lemma~\ref{lemma:exists-gfp}. We chose to restrict the verification
to $R=2^\lambda$ because 
this is
empirically
the hardest case. To obtain $\Delta(\lambda, 2^\lambda,
\technicalS\lambda 2^\lambda)$,
we used a simple primality proof algorithm based on
Pocklington's theorem, in Las Vegas
probabilistic time. The result of our experiments is given in
Table~\ref{tab:empirical-heuristic}.

Hypothesis~\ref{conj:second} is in fact stronger than what would be
strictly necessary to reach the asymptotic complexity we claim in this
article. 
Proposition~\ref{prop:estimate-is-big} led us to choose $\alpha$ as a
polynomial of degree at least two, and our particular choice
$\alpha=\technicalS\lambda$ has the advantage that the data in
Table~\ref{tab:empirical-heuristic} has no corner cases for small values
of $\lambda$ (in particular for $\lambda=3$).
\smallskip

\begin{table}
    \begin{center}

        $$\begin{array}{c|c|c|c}
            \lambda & \text{Candidates} & \text{Primes} & \text{Estimate}\\
            \hline
            1 & 0 & 0 & 0 \\
            2 & 9 & 3 & 5 \\
            3 & 58 & 1 & 8 \\
            4 & 248 & 24 & 22 \\
            5 & 878 & 31 & 30 \\
            6 & 2789 & 57 & 45 \\
        \end{array}\qquad\begin{array}{c|c|c|c}
                        \lambda &
                        \text{Candidates} &
                        \text{Primes} &
                        \text{Estimate}\\ 
                                    \hline 
            7 & 8233 & 42 & 46 \\
            8 & 2.3e4 & 126 & 138 \\
            9 & 6.2e4 & 184 & 170 \\
            10 & 1.6e5 & 224 & 218 \\
            11 & 4.1e5 & 227 & 230 \\
            12 & 1.0e6 & \geq307 & 312\\
        \end{array}$$
        \caption{%
            \label{tab:empirical-heuristic}
        Number of generalized
        Fermat primes $r^{2^{\lambda}}+1$ with
        $r\in\closedopeninterval{R,\technicalS\lambda R}$ with
        $R=2^\lambda$ (only even $r$ are counted as candidates), compared
        to
        the asymptotic estimate of Lemma~\ref{lemma:exists-gfp}.
        Hypothesis~\ref{conj:second} asserts that the third column is never
        zero for $\lambda\geq2$.}
    \end{center}
\end{table}

Throughout the rest of the article, Hypothesis~\ref{conj:second} is
tacitly assumed.

\subsection{Chains of generalized Fermat primes}\label{sec:chain-gfp}

Some generalized Fermat numbers, defined below, play a key role
in this article.
\begin{definition}[Admissible generalized Fermat number] \label{def:agfp}
    A generalized Fermat number 
    $p=P(r,\lambda)$ is called \emph{admissible} whenever $\lambda\geq
    4$  and $r$ is such that $2^\lambda\leq r<\technicalLexp\lambda$.
\end{definition}
Definition~\ref{def:agfp} captures the primes whose existence is asserted
by Hypothesis~\ref{conj:second} (it is easy to observe that these are
admissible when $\lambda\geq4$), as well as generalized Fermat
numbers that are subject to the same bounds.


The following proposition shows how from admissible generalized Fermat
numbers (not necessarily prime), we can build smaller
generalized Fermat primes. For large enough inputs, these smaller primes
are in turn admissible, so that this construction can be used another
time.

\begin{proposition}
    \label{prop:agfp-chain}
    Let $\lambda\geq4$, and let
    $p=P(r,\lambda)=\gfprime{r}{\lambda}$ be an admissible generalized
    Fermat number. A smaller generalized Fermat prime denoted
    $\smallerprime(p)$ and an integer $\batchsize(p)$ are defined as
    follows.

    Let $\lambda'=\left\lceil \lg\lg\lg p \right\rceil$.  Let
    $\phi(k)=2^{k+1}\lg r+\lambda-k$. There exists a power of two~$\beta$
    such that the following conditions hold:
    \begin{enumerate}
            \def\theenumi{{\itshape\roman{enumi}\/}}
        \item $0\leq\lg\beta<\lambda'$,
        \item $\lambda'2^{\lambda'}\leq\phi(\lg\beta)\leq2\lambda'2^{\lambda'}$,
        \item Given $R'=2^{\phi(\lg\beta)/2^{\lambda'}}$, there exists an
            integer $r'\in \closedopeninterval{R',\technicalS{\lambda'}R'}$ such that
            $p'=P(r',\lambda')=\gfprime{r'}{\lambda'}$ is a generalized
            Fermat prime.
    \end{enumerate}
    Given $\beta$ and $p'$ as above, we let $\smallerprime(p)=p'$ and
    $\batchsize(p)=\beta$.
    Furthermore, if $\lambda'\geq4$, then $p'$ is
    admissible too.
\end{proposition}

    In anticipation for the proof of Proposition~\ref{prop:agfp-chain},
    we prove the following bounds.
    \begin{lemma}
        \label{lemma:bound-lambda}
        Let $\lambda$ and $\lambda'$ be as in
        Proposition~\ref{prop:agfp-chain}. We have
        $$\lg(\lambda+\lg\lambda)\leq\lambda'<3\lg\lambda-1<\lambda.$$
    \end{lemma}
    \begin{proof}
    Since $p$ is admissible, we have
    $$2^{\lambda'}\geq\lg(2^\lambda\lg
        r)\geq\lambda+\lg\lg r\geq\lambda+\lg\lambda.$$
    In the other direction, the condition on $p$ being admissible gives
        the following uniform bound on $\lambda'$ (we first bound $p$ by
        $2r^{2^\lambda}$):
    $$\lambda'\leq
        1+\lg\lg(1+2^\lambda\technicalL\lambda).$$
    An unilluminating calculation shows that this right hand side is
        indeed bounded by $3\lg\lambda-1$ for all $\lambda\geq3$, and
        then by $\lambda$ for all $\lambda\geq4$.
%
%
    \end{proof}
    The lower bound given by Lemma~\ref{lemma:bound-lambda} is most useful now, and gives in fact the
    correct order of magnitude for $\lambda'$.  The upper bound is much
    coarser and will be used in §\ref{sec:solve-rec}.
    Possible values for $\lambda'$ are given in
    Table~\ref{tab:lambda-lambdaprime}. In particular,
    $\lambda\geq4$
    implies $\lambda'\geq3$.

    \begin{table}
        \begin{center}
            \begin{minipage}{.25\textwidth}
            $$\begin{array}{c|c} \lambda & \lambda'\\ \hline
                3 \leq\lambda\leq 4 & 3\\
                \lambda=5 & 3, 4\\
                6 \leq\lambda\leq 11 & 4\\
            \end{array}$$
            \end{minipage}\hfil
            \begin{minipage}{.25\textwidth}
            $$\begin{array}{c|c} \lambda & \lambda'\\ \hline
                \lambda= 12 & 4,5\\
                13 \leq\lambda\leq 26 &5\\
                \lambda= 27 & 5, 6\\
            \end{array}$$
            \end{minipage}\hfil
            \begin{minipage}{.25\textwidth}
            $$\begin{array}{c|c}
                \lambda & \lambda'\\ \hline
                28 \leq\lambda\leq 56 & 6\\
                57 \leq\lambda\leq 58 & 6,7\\
                59\leq\lambda & \geq 7
            \end{array}$$
            \end{minipage}
            \caption{%
                \label{tab:lambda-lambdaprime}
                Possible values for
            $\lambda'=\left\lceil\lg\lg\lg p\right\rceil$ for
            $p=P(r,\lambda)$ an admissible generalized
            Fermat number, using the bounds
            $\lg(\lambda+\lg\lambda)\leq\lambda'\leq
        1+\lg\lg(1+2^\lambda\technicalL\lambda
            )$.}
        \end{center}
    \end{table}

\begin{proof}[Proof of Proposition~\ref{prop:agfp-chain}]
    The function $\phi$ is easily seen to satisfy $\phi(k)\leq2\phi(k-1)$
    for any integer $k\leq\lambda+2$. As a consequence, the intervals
    $[\phi(k),2\phi(k)]$, for $k$ ranging from $0$ to $\lambda'-1$, form a
    covering of the interval $[\phi(0),\phi(\lambda')]$.

    We prove $\phi(0)\leq 2\lambda'2^{\lambda'}\leq\phi(\lambda')$, which
    will directly entail that $2\lambda'2^{\lambda'}$ is within one of the above
    intervals that form a covering.

    The bound $2\lambda'2^{\lambda'}\leq\phi(\lambda')$ is a consequence
    of
    $\lambda\geq\lambda'$: $$\phi(\lambda')\geq 2^{\lambda'+1}\lg
    r\geq2^{\lambda'+1}\lambda\geq2^{\lambda'+1}\lambda'.$$

    The proof that $2\lambda'2^{\lambda'}\geq\phi(0)$ is based on
    calculus. Lower and upper bounds for $2\lambda'2^{\lambda'}$ and $\phi(0)$
    are
    \begin{align*}
        2\lambda'2^{\lambda'}&
    \geq2(\lambda+\lg\lambda)\lg(\lambda+\lg\lambda)
        \geq(\lambda+\lg\lambda)\lg(36)
        ,\\
	    \phi(0)&\leq\lambda+2\technicalL\lambda=5(\lambda+\lg\lambda).
    \end{align*}
%
%
%

    We have proved that there exists an integer $k$ such that $0\leq
    k<\lambda'$, and that $\phi(k)\leq2\lambda'2^{\lambda'}\leq2\phi(k)$.
    Let $\beta=2^k$, so that 
    ({\it i\/}) holds.
    We have that
    $$\lambda'\leq\frac{\phi(\lg\beta)}{2^{\lambda'}}\leq2\lambda'.$$
    This implies ({\it ii\/}). Finally,
    $R'=2^{\frac{\phi(\lg\beta)}{2^{\lambda'}}}$ is such that
    $2^{\lambda'}\leq R'\leq2^{2\lambda'}$.  
    Hypothesis~\ref{conj:second} then implies ({\it iii\/}),
    and concludes the proof. Admissibility of $p'$ follows from
    Definition~\ref{def:agfp}.
\end{proof}
The following technical lemma provides useful bounds
for $p'=\smallerprime(p)$.
\begin{lemma}
    \label{lemma:bound-logprime}
    Let $p=P(r,\lambda)$ be as in Proposition~\ref{prop:agfp-chain}. Let
    $\beta=\batchsize(p)$ and $p'=\smallerprime(p)$. We
    have
    \begin{enumerate}
            \def\theenumi{{\itshape\roman{enumi}\/}}
        \item
            $1\leq\frac{\lg p'}{2\beta\lg r}\leq\min(
            1+\frac{4\lg\lambda'}{\lambda'-1}, \frac72)$.
            In particular, $\frac{\lg p'}{\beta\lg r}=2+o(1)$.
        \item $\lambda'+\lg\lambda'\leq\lg\lg
            p'\leq\lambda'+\lg\lambda'+2$.
    \end{enumerate}
\end{lemma}
\begin{proof}
    We follow the notations of Proposition~\ref{prop:agfp-chain}.
    The lower bound in ({\it i\/}) is easy:
    $$\lg p' \geq 2^{\lambda'}\lg R' \geq \phi(\lg\beta) \geq 2\beta\lg
    r.$$
    The upper bound requires more work.
    On the
    one hand, Lemma~\ref{lemma:bound-lambda} gives $2^{\lambda'}>\lambda$,
    whence
    \begin{gather*}
        2\beta\lg r + 2^{\lambda'}
        \geq\phi(\lg\beta)\geq\lambda'2^{\lambda'}\\
        2\beta\lg r\geq{{(\lambda'-1)} 2^{\lambda'}}
        .
    \end{gather*}
    And on the other hand, we can bound $\lg p'$ as follows.
    \begin{align}
        \notag \lg p' &= \lg((p'-1)+1) = \lg(p'-1) + \lg(1+1/(p'-1))\\
        &\leq 2^{\lambda'}\lg r' + 1
        \notag \leq 2^{\lambda'}\lg R' + 2^{\lambda'}\lg(\technicalS{\lambda'}) +
        1\\
        \label{ineq:loglogprime}
        &\leq \phi(\lg\beta) + 2^{\lambda'}\lg(\technicalS{\lambda'}) +
        1
        \intertext{by the definition of $R'$. Using now
        Lemma~\ref{lemma:bound-lambda} and
        $2^{\lambda'}\geq\lambda+\lg\lambda\geq\lambda+2$ we have}
        \notag 
        \lg p'
        &\leq 2\beta\lg r + (2^{\lambda'}-2) +
        2^{\lambda'}\lg(\technicalS{\lambda'}) +
                1\\
        \notag &\leq 2\beta\lg r+2^{\lambda'}\cdot
        \min(4\lg\lambda',5(\lambda'-1)/2)\rlap{\ \ since
        $\lambda'\geq3$.}
    \end{align}
%
%
    The upper bound on the last line is obtained by calculus.
    We have thus proved ({\it i\/}).
%

    The lower bound in statement ({\it ii\/}) is trivial. The upper bound is
    derived from inequality~\eqref{ineq:loglogprime} above.
    By ({\it ii\/})
    in
    Proposition~\ref{prop:agfp-chain}, we have
    $\frac{\phi(\lg\beta)}{2^{\lambda'}}\leq 2\lambda'$, whence
    \begin{align*}
        \lg p'
        &\leq 2\lambda'2^{\lambda'}+
        2^{\lambda'}\lg(\technicalS{\lambda'}) + 1.\\
        \lg\lg p'&\leq
        \lg\left(1+2^{\lambda'}(2\lambda'+\lg(\technicalS{\lambda'}))\right)\\
        &\leq \lambda'+\lg\lambda'+2\ \text{since $\lambda'\geq3$.}
    \end{align*}
        Again, this last upper bound is verified by calculus.
%
%
\end{proof}

\section{Two new algorithms}\label{sec:algo}

We now see how we can design an asymptotically fast integer
multiplication algorithm that uses rings of integers modulo generalized
Fermat primes.

Throughout this section, our preferred representation for elements of a
ring \cR\ of integers modulo a generalized Fermat number $p=P(r,\lambda)$
is the representation \emph{in radix $r$}. Namely, $a\in\cR$ is
represented as a $2^\lambda$-uple $(a_0,\ldots,a_{2^\lambda-1})$ such
that $a=\sum_{j<2^\lambda}a_jr^j$ and $0\leq a_j<r$. This representation does not cover the
case $a=-1$, and we need an \emph{ad hoc} exceptional representation for
this case (possible representation
choices are plenty -- one extra bit is enough). Conversions between binary
representation and radix $r$ representation can be done in linear time
when $r$ is a power of two, but we also need to deal with the general
case. 
Recursive base conversion algorithms
(see~\cite[§1.7.2]{BrZim:2010}), do this in quasi-linear time
$O(\lambda\sM(\log p))$ (this holds both for ways, both
\emph{to} and \emph{from} representation in radix $r$). Additions and
subtractions in
\cR\ using this representation are linear. This section is
concerned
with the complexity of multiplication in \cR. We denote this cost by
$\sM_{\cR}$.

\subsection{Preliminaries: transforms}
\label{subsec:algo:transforms}
The following definition extends concepts defined in
Proposition~\ref{prop:agfp-chain} and defines useful data for our
algorithms.
\begin{definition}[$\smallerring(\cR)$]
    \label{def:smallerring}
Let $\lambda\geq4$.
Let $p=P(r,\lambda)=\gfprime r\lambda$ be an
admissible generalized Fermat number, and let $\cR=\ZnZ p$. Following
    Proposition~\ref{prop:agfp-chain} we let
    $\smallerring(\cR)$ be the triple $(\cR',N',\omega')$
    defined as follows:
\begin{itemize}
    \item $\cR'=\ZnZ{p'}$, with $p'=P(r',\lambda')=\smallerprime(p)$.
    \item $N'=2^\lambda/\batchsize(p)$.  ($N'$ is a power of two.)
    \item $\omega'$ is a primitive $2N'$-th root of unity in $\cR'$.
\end{itemize}
\end{definition}

For the root $\omega'$ to be well defined above, we need
the following property.

\begin{lemma}
\label{lemma:root-in-smallerring}
    Using the notations above, $\cR'$ has a primitive $2N'$-th root of unity.
\end{lemma}
\begin{proof}
    Notice first that $\lambda'\geq3$ so that
    $p'\geq 2^{3\cdot 2^3}$, and that $p'=\gfprime{r'}{\lambda'}$ is
    prime, so
    that in particular $r'$ is even.
    For $2N'$ to divide $p'-1$, it
    suffices to check that $2N'$ divides $2^{2^{\lambda'}}$. We have
    \begin{align*}
        \lg(2N') &= \lambda+1-\lg\beta\\
        2^{\lambda'} &\geq \lambda + \lg\lg r,
    \end{align*}
    so that it is sufficient to check that $\lg\lg r\geq 1$, which holds
    as soon as $\lambda\geq2$.
\end{proof}

The algorithms described in the remainder of this section all assume that the sequences of rings and
auxiliary data defined by Definition~\ref{def:smallerring} are computed
in advance, for all levels of the recursion. We assume that a tape of our
Turing machine is devoted to that data, stored one level after another.
The size of the data $\smallerring(\cR')$ is clearly $O(\log p')$.

\subsection{New algorithms}
\label{subsec:new-algorithms}

We now describe two new algorithms that are dependent on each other. Both
aim at computing products of elements of \cR.
\begin{itemize}
    \item One algorithm that computes ``transforms'' of elements of \cR.
        Internally, this algorithm multiplies elements of $\cR'$.
    \item One algorithm that multiplies elements of $\cR$. This
        algorithm uses the transforms computed by the previous algorithm.
\end{itemize}

We begin with \algorithmref{algo:FFTR}, which computes transforms. We can
state it thanks to
Lemma~\ref{lemma:root-in-smallerring}.

\begin{myalgorithm}{TransformR}
    \caption{Transform $\cT_\cR(a)$ of $a\in\cR=\ZnZ p$, with $p=\gfprime r\lambda$
    admissible (not necessarily prime), $\lambda\geq4$.
    (Algorithm without
    precomputations.)
    }
\algorithmlabel{algo:FFTR}
    \begin{algorithmic}[1]
	\Function{\currentalgorithm}{$a$}
        \Input{$a\in\cR$, represented in radix $r$.}
        \Output{$\cT_\cR(a)$, a vector of $N'$ elements of $\cR'$,
        represented in radix $r'$}
        \State Let $\beta=\batchsize(p)$, and
        $(\cR',N',\omega')=\smallerring(\cR)$.
    \State%
        \label{algline:FFTR:radixr1}%
        \label{algline:FFTR:Atilde}%
        Let $\tilde A(X)\in\bZ[X]$ with positive coefficients below
        $r^\beta$ be such that $\tilde A(r^\beta)=a$;
    \State
        \label{algline:FFTR:A}%
        Map $\tilde{A}$ to $A\in \cR'[X]/(X^{N'}+1)$.
    \State
        \label{algline:FFTR:radixr2}%
        Rewrite coefficients of $A$ in radix $r'$.
    \State
        \label{algline:FFTR:HDFT}%
        \Return {$
        \HDFT_{N',\omega'}(A)=\textproc{LargeRadixFFT}(N',{\omega'}^2,2^{\lambda'+1},A(\omega'X))$.}
	\EndFunction
\end{algorithmic}
\end{myalgorithm}

Our complexity analysis will need to reason on the set of transforms
of roots of unity that are used by
Algorithm~\ref{algo:FFTR}. We define it
as follows:
\begin{definition}[$\cW(\cR)$, vector of precomputed transforms useful
    for $\cT_\cR$]
    \label{def:WR}
    Fix notations as in Definition~\ref{def:smallerring}. We let
    $\cW(\cR)$ denote the vector defined as:
    $$\mathcal{W}(\cR)=\{\cT_{\cR'}({\omega'}^{2i+1}),\
        i\in\closedintervalZ{0,\frac{N'}{2^{\lambda'+1}}-1}\}$$
    where $\cT_{\cR'}$ is defined as in Algorithm~\ref{algo:FFTR} (albeit using
    $\cR'$ as an input ring).
\end{definition}

\subsubsection*{Complexity of Algorithm~\ref{algo:FFTR}, with or without
precomputations}
We define the following costs. The analysis of $\sM_\cR$ and $\sM'_\cR$
will be done in §\ref{sec:mulR}.
\begin{itemize}
    \item $\sM_\cR$:~cost of multiplying $a\in\cR$ by $b\in\cR$, with no auxiliary inputs.
    \item $\sM'_\cR$:~cost of the same computation, with $\cT_\cR(b)$
        known.
    \item $\sT_\cR$:~cost of computing $\cT_\cR$ with
        Algorithm~\ref{algo:FFTR}.
    \item $\sT'_\cR$:~cost of computing $\cT_\cR$ with
        Algorithm~\ref{algo:FFTR}, aided with the auxiliary knowledge of
        $\cW(\cR)$.
    \item $\sW_\cR$:~cost of computing $\cW(\cR)$.
\end{itemize}

We begin with $\cT_\cR$.
Algorithm~\ref{algo:FFTR} uses base conversions on
lines~\ref{algline:FFTR:radixr1} and~\ref{algline:FFTR:radixr2}. Both
operations perform $N'=2^\lambda/\beta$ conversions, and the respective
costs per conversion in each case are $O(\log\beta\cdot\sM(\beta\log r))$
and $O(\lambda'\cdot\sM(\beta\log r))$ (in these complexity estimates,
$\sM(n)$ can be taken as the complexity obtained for mutiplying integers
by the Schönhage-Strassen algorithm, for example).  By
Proposition~\ref{prop:agfp-chain} we have $\log\beta\leq\lambda'$, and by
Lemma~\ref{lemma:bound-logprime} we have $\log p'=\Theta(\beta\log r)$,
so that the overall base conversion costs in Algorithm~\ref{algo:FFTR}
can be expressed as $O(N'\lambda'\cdot\sM(\log p'))$.

The computation of the $\HDFT$ on line~\ref{algline:FFTR:HDFT} of
Algorithm~\ref{algo:FFTR} involve $N'\log N'$ multiplication by roots of
unity in $\cR'$, of which only $(E(N')+N')$ exceed a linear cost
(using the notation of §\ref{subsec:polring}).
We
have
$$\sT_\cR =
(E(N')+N')\sM_{\cR'} + 
O\left(
N'\log N'\log p' +
N'\lambda'\cdot\sM(\log
p')\right).$$

We now turn to the analysis of $\sT'_\cR$.
If the vector $\cW(\cR)=\{\cT_{\cR'}(\omega^{2i+1}),\
i\in\closedintervalZ{0,\frac{N'}{2^{\lambda'+1}}-1}\}$ is known, then the
computation of $\cT_\cR$ can be
done a bit faster: the $(E(N')+N')$ ``expensive'' multiplications by roots of unity in $\cR'$ do not
need to recompute the transforms of the roots. They may thus use a
somewhat faster algorithm for multiplication in $\cR'$. We defined above
its cost as
$\sM'_{\cR}$, and we have:
$$\sT'_\cR = (E(N')+N')\sM'_{\cR'} +
O\left(
N'\log N'\log p' +
N'\lambda'\cdot\sM(\log
p')\right).$$ 

Finally, we give the cost $\sW_\cR$ of computing $\cW(\cR)$. 
Here, we do \emph{not} recursively use $\cW({\cR'})$ to compute the
different elements. We do however use the knowledge of the root of unity
$\omega'$ (it belongs to the precomputed data $\smallerring(\cR)$). To
compute $\sW_\cR$, we first compute $\cT_{\cR'}(\omega')$ and
$\cT_{\cR'}({\omega'}^2)$ , which cost
$2\sT_{\cR'}$. Then we do successive pointwise multiplications by the
vector $\cT_{\cR'}({\omega'}^2)$ to obtain the transforms of the other roots.
For each of the $N'/2^{\lambda'+1}-1$ transforms to be inferred this way, we need $N''$
multiplications in $\cR''$, where we temporarily set
$(\cR'',N'',\omega'')=\smallerring(\cR')$. Therefore we have
$$\sW_\cR \leq 2\sT_{\cR'} + (N'/2^{\lambda'+1}-1)N''\sM_{\cR''}\leq N'\sT_{\cR'}.$$

Without further detail, we also claim that the inverse transform
$\cT^{-1}_\cR$ can be computed with the same cost as $\cT_\cR$.

\subsection{Multiplication modulo generalized Fermat numbers}\label{sec:mulR}

\begin{myalgorithm}{MulR}
    \caption{Multiplication in $\cR=\ZnZ p$, with $p=\gfprime r\lambda$
    admissible, $\lambda\geq4$.\goodbreak
    $p$ is not necessarily prime.}
\algorithmlabel{algo:mulR}
    \begin{algorithmic}[1]
        \Statex We use the notations $\cT_\cR$, $\cW_\cR$ as in
        §\ref{subsec:new-algorithms}.
        \Function{\currentalgorithm}{$a$,$T_\cR(b)$}
        \Input{$a\in\cR$,
        represented
        in radix $r$; $\cT_\cR(b)$ for some $b\in\cR$.}
	\Output{$a\cdot b \mod p$, represented in radix $r$}
        \State Let $\beta=\batchsize(p)$, and
        $(\cR',N',\omega')=\smallerring(\cR)$.
    \State Compute $W=\cW(\cR)$ using Algorithm
        \textproc{TransformR}.
    \State Compute $\cT_\cR(a)$ using Algorithm
        \textproc{TransformR} and $W$ as auxiliary data.
        \label{algline:mulR:radixr1}%
        \label{algline:mulR:ABtilde}%
    \State
        \label{algline:mulR:pointwise}%
        Compute
        $\gamma=\cT_\cR(a)*\cT_\cR(b)$
        \Comment pointwise products of elements of $\cR'$.
        \State Compute $c=\cT_\cR^{-1}(\gamma)$ as follows:
    \BeginBlock
        \State
            $C\gets\operatorname{Half-IFT}_{N',\omega'}(\gamma)\in
            \cR'[X]/(X^{N'}+1)$ using $W$ as auxiliary data.
        \State
            \label{algline:mulR:Ctilde}
            Lift $C$ to $\tilde C\in\bZ[X]$ as follows:
        \BeginBlock
            \For{$i \in \closedintervalZ{0,N'-1}$}
                \State
                    Lift coefficient of degree $i$ to
                    $\closedopenintervalZ{-(N'-1-i){r^{2\beta}},(i+1){r^{2\beta}}}$.
                \EndFor
            \label{algline:mulR:Ctilde-end}
        \EndBlock
        \State
            \label{algline:mulR:radixr2}
            Rewrite coefficients of $\tilde C$ as signed integers in
            radix $r$.
        \State Compute $\tilde C(r^\beta)=c$.  \Comment The result is defined
            modulo $(r^\beta)^{N'}+1=p$.
    \EndBlock
    \State \Return $c$
	\EndFunction
\end{algorithmic}
\end{myalgorithm}

Using \algorithmref{algo:FFTR}, we can now state
\algorithmref{algo:mulR}. Its
validity depends on the following lemma:

\begin{lemma}\label{lemma:mulR-checks}
    Let notations be as in Algorithm~\ref{algo:mulR}.  Let $\tilde
    A,\tilde B$ be 
    polynomials in $\bZ[X]$ of degree less than $N'$ and with
    positive coefficients below $r^\beta$ such that, $A$ and $B$ being
    their respective images in $\cR'[X]/(X^{N'}+1)$, we have
    $\cT_\cR(a)=\HDFT_{N',\omega'}(A)$ and
    $\cT_\cR(b)=\HDFT_{N',\omega'}(B)$ on
    line~\ref{algline:mulR:pointwise} of Algorithm~\ref{algo:mulR}.
    \begin{enumerate}
        \def\theenumi{{\itshape\roman{enumi}\/}}
        \item Both $\tilde A$ and $\tilde B$ are uniquely defined from
            $\cT_\cR(a)$ and $\cT_\cR(b)$.
        \item
            The polynomial $\tilde C$ is equal to $\tilde A\cdot \tilde
            B\bmod X^{N'}+1$.
        \item $c$ is equal to $ab\bmod p$.
    \end{enumerate}
\end{lemma}
\begin{proof}
    We prove ({\it i\/}) for $\cT_\cR(a)$, the same reasoning holds for
    $\cT_\cR(b)$.
    The polynomial $A\in\cR'[X]/(X^{N'}+1)$ is uniquely defined because $\HDFT$ is an
    isomorphism. Now since $\cT_\cR(a)$ is
    computed from an element
    $a$ of \cR, line~\ref{algline:FFTR:Atilde} of
    Algorithm~\ref{algo:FFTR} has unambiguously computed a polynomial
    $\tilde A$, which meets the conditions. Since there is a unique lift
    of $A$ to $\bZ[X]$ that has degree less than $N'$ and positive
    coefficients below $p'$, this lift is then necessarily the same as $\tilde
    A$.

    Statement ({\it ii\/}) holds modulo $p'$ by construction, but we must
    make sure that the lift on
    lines~\ref{algline:mulR:Ctilde}-\ref{algline:mulR:Ctilde-end} of
    Algorithm~\ref{algo:mulR} computes the correct product over
    the integers.
    To do so, we compute a bound for the coefficients of the
    product $\tilde A\cdot \tilde B\bmod X^{N'}+1$.
    Both operands have at most $N'$
    coefficients. The coefficient of degree $i$ of their
    product modulo $X^{N'}+1$ lies within the interval
    $\closedopenintervalZ{-(N'-1-i)(r^\beta)^2,(i+1)(r^\beta)^2}$
    (actually with the lower endpoint open for $i<N'-1$), which has width
    $N'(r^\beta)^2$. The base 2 logarithm of this latter value is
    $2\beta\lg r+\lambda-\lg\beta=\phi(\lg\beta)$, following the notation of
    Proposition~\ref{prop:agfp-chain}. Now again following notations of
    Proposition~\ref{prop:agfp-chain}, we have $p'\geq
    {R'}^{2^{\lambda'}}\geq 2^{\phi(\lg\beta)}\geq N'(r^\beta)^2$.
    Thus, the coefficient $c_i$ of degree $i$ of $\tilde A\cdot \tilde B\bmod
    X^{N'}+1$ is lifted to a 
    unique signed representative modulo $p'$ on
    line~\ref{algline:mulR:Ctilde}.
    This proves the claim.\footnote{On
    lines~\ref{algline:mulR:Ctilde}-\ref{algline:mulR:Ctilde-end} of
    Algorithm~\ref{algo:mulR}, 
    intervals
    depend on the degree so that we can do without a needlessly coarse
    lower bound $2N'(r^\beta)^2\leq p'$.
    It would be possible to adjust the definition of $\phi$
    in Proposition~\ref{prop:agfp-chain}, as well as the corresponding
    proofs,
    so that that coarser
    inequality holds.}

    Statement ({\it iii\/}) follows: by ({\it ii\/}), we have that $\tilde
    C = \tilde A\cdot \tilde B\mod X^{N'}+1$.  By evaluating at $r^\beta$,
    we obtain the result $c=ab$ modulo $(r^\beta)^{N'}+1=p$.
\end{proof}
\subsubsection*{Complexity analysis of Algorithm~\ref{algo:mulR}} 

We first mention that the relative costs of multiplications and transforms, with or without
precomputations, satisfy the following equations.
\begin{eqnarray*}
2\sT'_\cR\leq\sM'_\cR\leq\sM_\cR\leq\sM'_\cR+\sT'_\cR\leq\frac32\sM'_\cR
    &\text{and}\quad
\sT_\cR\leq\sM_\cR.
\end{eqnarray*}
(To get $\sM_\cR\leq\sM'_\cR+\sT'_\cR$, it suffices to \emph{first} 
compute $\cW(\cR)$, and then $\cT_\cR(b)$.)

On line~\ref{algline:mulR:Ctilde},
Algorithm~\ref{algo:mulR} converts
between
representation in radix $r'$
and
binary representation. On line~\ref{algline:mulR:radixr2} the conversion
is between binary representation and representation in radix $r$. As with
Algorithm~\ref{algo:FFTR}, we can do this in time
$O(N'\lambda'\cdot\sM(\log p'))$. Pointwise products, on
line~\ref{algline:mulR:pointwise}, use a variation of Algorithm~\ref{algo:mulR}, where there is no
auxiliary input, recursively
(thus exploiting the fact that the coefficients of $\cT_\cR(a)$ and
$\cT_\cR(b)$ are represented in radix $r'$). And last but not least, the most important aspect of the
complexity of Algorithm~\ref{algo:mulR} is that since we compute
$\cW(\cR)$, the transforms $\cT_\cR(a)$ and $\cT_\cR^{-1}(\gamma)$ can
take advantage of it.
We thus have:
\begin{align*}
    \sM'_\cR &= \sW_{\cR}
    + 2\sT'_{\cR}
    + N'\sM_{\cR'}
    + O(N'\lambda'\cdot\sM(\log p'))
    + O(\log p)
    .
    \intertext{We now use the various expressions obtained in
    §\ref{subsec:new-algorithms} to rewrite this. We use the coarse
    bounds
    $\sT_{\cR'}\leq \sM_{\cR'}\leq\frac32\sM'_{\cR'}$.
    We have}
    \sM'_\cR &\leq N'\sM_{\cR'}
    + 2E(N')\sM'_{\cR'} + 2N'\sM'_{\cR'}
    + N'\sM_{\cR'}\\&\qquad
    + O(N'\cdot\lambda'\cdot\sM(\log p'))
    + O(N'\cdot\log N'\cdot\log p')
    + O(\log p)\\
    &\leq
    5N'\sM'_{\cR'}
    + 2E(N')\sM'_{\cR'}\\&\qquad
    + O(N'\cdot\lambda'\cdot\sM(\log p'))
    + O(N'\cdot\log N'\cdot\log p')
    + O(\log p)\\
    &\leq
    2N'\cdot\left(3+\log_{2^{\lambda'+1}}N'\right)\cdot \sM'_{\cR'}\\&\qquad
    + O(N'\cdot\lambda'\cdot\sM(\log p'))
    + O(N'\cdot\log N'\cdot\log p')
    + O(\log p)
\end{align*}
where we used $E(N')\leq
N'\log_{2^{\lambda'+1}}N'$ and $5/2<3$. Algorithm~\ref{algo:mulR} also
needs to move the head of tape of precomputed data by the size of the current
data $\smallerring(\cR)$. The corresponding overhead $O(\log p')$ is easily subsumed within
the lower-order terms above.


\subsection{Multiplication in \bZ\ using multiplication in \cR}\label{sec:mulZ} 
We can build on Algorithm~\ref{algo:mulR} to obtain an
integer multiplication algorithm for $n$-bit integers $a$ and $b$.

Note however that we avoid the following simple approach because it does
not work complexity-wise: we do not multiply $a$ and $b$ by
considering them as elements of \ZnZ p for $p$ an admissible generalized
Fermat number such that $p\geq 2^{2n}$. There are two
reasons for that. First, doing so for $p$ an admissible generalized
Fermat \emph{prime} is out of question:
unless we consider that $p$ is given beforehand, computing it is
likely to be more expensive than computing a product of bit length $\lg p$,
and would therefore appear dominant, maybe prohibitive even for a
precomputation. Fortunately, \algorithmref{algo:mulR} does not require
that $p$ be prime, and therefore this difficulty can easily be
circumvented.
    For example we may select
    $\lambda$ such that $\lambda 2^{\lambda}\geq 2n$, and then set
    $p=P(2^{\lambda},{\lambda})$.
The second issue is harder to deal with: in the ring
$\cR'$ used by Algorithm~\ref{algo:mulR}, we need to find $2N'$-th roots
of unity, and for this we need a quadratic nonresidue
in $\cR'$ (which generates the $2$-Sylow subgroup of $\cR'$). Alas, if our first (non-prime) modulus $p$ is such
that $\lg p\geq 2n$, then in
Proposition~\ref{prop:agfp-chain} we have $\lambda'=\lceil\lg\lg\lg
p\rceil\geq\lg\lg n$, so that the upper bound on $\lg p'$ that we obtain
from Lemma~\ref{lemma:bound-logprime} is at least as large as $\lg
n\cdot\lg\lg n$. If we can use only deterministic exponential-time
algorithms to search for a quadratic nonresidue in $\cR'$, then the
complexity of this search exceeds the overall complexity of integer
multiplication.

Similar (but subtly different) issues were already encountered by Harvey, van der Hoeven and
Lecerf.
The workarounds proposed
in~\cite[§8]{HavdHLe16} 
also apply here.
\begin{itemize}
    \item Either we assume the generalized Riemann hypothesis, in which
        case a quadratic nonresidue in $\cR'$ can
        be found in polynomial time.
    \item Or we do the top-level multiplication with one round of Fürer's
        algorithm.
Multiplication in the ring
\quotient{\bC[X]}{X^{2^\lambda}+1} that is used by
\algorithmref{algo:multfurer}
        reduces to multiplication of integers of
bit length $n_0=O((\log n)^2)$, with $n$ denoting the bit length of the
integers $a$ and $b$ (see Equation~\eqref{eq:cplx-furer}). These integers are
then multiplied by
        \algorithmref{algo:mulR}, for a suitable modulus $p_0$ (not
        necessarily prime).
\end{itemize}

The latter strategy is given by \algorithmref{algo:mulZ}. Note that since
we build upon Algorithm~\ref{algo:multfurer}, we force the bit length $n$
to be rounded up to a power of two.

\begin{myalgorithm}{MulZ}
    \caption{%
\algorithmlabel{algo:mulZ}
        Multiplication of integers in $\bZ$}
    \begin{algorithmic}[1]
        \Input{$a$, $b$ two positive $n$-bit integers, $n$ being a power
        of two.}
        \Output{$c=a\cdot b$}
	\Function{\currentalgorithm}{$a$,$b$}
        \State
        Let $n_0$ be such that all internal multiplications in
        $\textproc{FurerComplexMul}(\cdot,\cdot,n)$ may be done by multiplying
        two $n_0$-bit integers. (As per the analysis of
        \textproc{FurerComplexMul}, we have $n_0=O((\lg n)^2)$.)
        \State
        Let $\lambda_0$ be the smallest integer such that $2n_0\leq
        \lambda_02^{\lambda_0}$.
        \State Let $p_0=P(2^{\lambda_0},{\lambda_0})=2^{\lambda_02^{\lambda_0}}+1$.
        \State
        \Return $c=\textproc{FurerComplexMul}(a,b,n)$, where 
        all internal multiplications 
        are done with
        \algorithmref{algo:mulR}, in the ring $\cR_0=\ZnZ{p_0}$.
	\EndFunction
\end{algorithmic}
\end{myalgorithm}

It is easy to see that $p_0$ in Algorithm~\ref{algo:mulZ} is an admissible generalized Fermat number.
As for the determination of \emph{prime} moduli as well as the
computation of primitive roots of unity of the desired order in the
recursive multiplication levels of Algorithm~\ref{algo:mulR}, we have
that $\lg(\smallerprime(p_0))$ is polynomial in $\lg\lg n$. This is
small enough so that simple algorithms are fit for the task of testing
$\smallerprime(p_0)$ for primality, as well as for finding primitive
roots. Thus the complete chain of precomputed triples
defined by $\smallerring$ in Definition~\ref{def:smallerring} can be
computed in advance and stored on an auxiliary tape of the Turing
machine, as suggested in §\ref{subsec:algo:transforms}.

The complexity of computing $n$-bit products with
\algorithmref{algo:mulZ}, which we denote by $\sM_{\text{new}}(n)$, can
be expressed as follows. The equation below is naturally very similar to 
Equation~\eqref{eq:cplx-furer}.
\begin{align*}
    \sM_{\text{new}}(n) &=
 N(3\lceil \log_{{2^{\lambda+1}}} N \rceil +1)\cdot \sM_{\cR_0} + O(N\log N \cdot {2^{\lambda}}\log n);\\
\end{align*}

\section{Solution of the recursive complexity equations}\label{sec:solve-rec}

\subsection{Summary of the recursive complexity equations}


In \algorithmref{algo:mulR}, multiplication in $\cR$
uses $(\cR',N',\omega')=\smallerring(\cR)$. In turn, multiplication in $\cR'$
may use $(\cR'',N'',\omega'')=\smallerring(\cR')$ if recursion is used again.
We define $(\cR_i)_{i\geq0}$ as well as $(N_i)_{i\geq1}$ and
$(\omega_i)_{i\geq1}$ by:
\begin{align*}
    \cR_0 &= \text{as in Algorithm~\ref{algo:mulZ}},\\
    (\cR_{i+1},N_{i+1},\omega_{i+1}) &= \smallerring(\cR_i)\ \text{for $i\geq0$}.
\end{align*}
Likewise, we let $p_i$ be such that $\cR_i=\ZnZ{p_i}$, for $i\geq0$.
Of course, since
Definition~\ref{def:smallerring} as well as Algorithms~\ref{algo:FFTR}
and~\ref{algo:mulR} are only valid for $\lambda_i\geq4$, only a finite
number of terms of the above sequences are defined for a given input size~$n$. Part of the work towards determining our final complexity will be to
determine this number of terms (the recursion depth).
We briefly recall the key equations for the complexity analysis:
\begin{align*}
    \sM'_{\cR_i}&\leq
    2N_{i+1}\cdot(3+\log_{2^{\lambda_{i+1}+1}}N_{i+1})\cdot \sM'_{\cR_{i+1}}\\&\qquad
    + O(N_{i+1}\cdot \lambda_{i+1}\cdot\sM(\log p_{i+1}))\\&\qquad
    + O(N_{i+1}\cdot\log N_{i+1}\cdot\log p_{i+1})\\&\qquad
    + O(\log p_i).\\
    \sM_{\cR_i}&\leq\frac32\sM'_{\cR_i}.\\
    \sM_{\text{new}}(n) &=
 N(3\lceil \log_{{2^{\lambda_0+1}}} N \rceil +1)\cdot \sM_{\cR_0} + O(N\log
    N \cdot {2^{\lambda_0}}\log n).
\end{align*}

We first prove the following that lemma bounds the transform length $N'$.
\begin{lemma}
    \label{lemma:bound-Nprime}
    Using the notations as above,
    we have
    $$N_{i+1}\leq\min\left(
    2
    \left(1+\frac{4\lg\lambda_{i+1}}{\lambda_{i+1}-1}\right),7
    \right)
    \cdot
    \frac{\lg p_i}{\lg p_{i+1}}
    .$$
\end{lemma}
\begin{proof}
    Let $\beta=\batchsize(p_i)$.
    We have  $2^{\lambda_i}\lg
    r_i\leq\lg p_i$, therefore $$N_{i+1}=\frac{2^{\lambda_i}}{\beta}
    \leq\frac{\lg p_i}{\beta\lg r_i}
    \leq2\frac{\lg p_i}{\lg p_{i+1}}\frac{\lg p_{i+1}}{2\beta\lg r_i}.
    $$
    Then 
    ({\it i\/}) in Lemma~\ref{lemma:bound-logprime} allows to conclude.
\end{proof}

The following result plays a central role in the asymptotic analysis.
\begin{proposition}
    \label{prop:cplx-mi}
    We keep the above notations. Let $i\geq0$ be such that $p_i$ is
    admissible.
    Let $\epsilon_{0,i}=\frac{4\lg\lambda_{i+1}}{\lambda_{i+1}}$,
    $\epsilon_{1,i} =
    \frac{8\lg\lambda_i}{\lambda_i}$, and
    $\epsilon_{2,i}=\frac{2+\lg\lambda_{i+1}}{\lambda_{i+1}}$. Let
    $m_i=\frac{\sM'_{\cR_i}}{\lg p_i\cdot\lg\lg p_i}$. We have
    $$m_i \leq4
    \cdot (1+\epsilon_{0,i})
    \cdot (1+\epsilon_{1,i})
    \cdot (1+\epsilon_{2,i})
    \cdot m_{i+1} + O(1).$$
\end{proposition}
\begin{proof}
We first bound the second and third lines in the equation for
$\sM'_{\cR_i}$, and compare them to $\log p_i\cdot\log\log p_i$.
The third line uses Lemma~\ref{lemma:bound-Nprime}. We have
    \begin{align*}
        N_{i+1}/\lg p_i&\leq 7/\lg p_{i+1}=O(1)\\
        \intertext{which obviously also implies $({\lg N_{i+1}})/({\lg\lg
        p_{i}}) = O(1)$. Then}
\frac{N_{i+1}\lg N_{i+1}\lg p_{i+1}}{\lg p_i\lg\lg p_i}
        &\leq 7\frac{\lg N_{i+1}}{\lg\lg p_{i}} = O(1).
    \end{align*}
For the second line, it
suffices to assume that $\sM(\log p_{i+1})$ is bounded by the complexity
of the Schönhage-Strassen algorithm. We have
\begin{align*}
    \frac{N_{i+1}\lambda_{i+1}\sM(\log p_{i+1})}{\lg p_i\lg\lg p_i}
    &\leq 7\frac{\lambda_{i+1}\lg\lg p_{i+1}\lg\lg\lg p_{i+1}}{\lg \lg
    p_i}=O(1).
\end{align*}
In the expression above, we obtain the upper bound by bounding the
    numerator by a polynomial in $\lambda_{i+1}$ (because $p_{i+1}$ is
    admissible), while the denominator is exponential in $\lambda_{i+1}$.

The most important calculation for the analysis is the comparison of the
first term of $\sM'_{\cR_i}$ with $\log p_i\cdot\log\log p_i$.
    Lemma~\ref{lemma:bound-Nprime} gives the bound
    $N_{i+1}\leq2(1+\epsilon_{0,i})\frac{\lg p_i}{\lg p_{i+1}}$, and we
    also have the coarse bound $\lg
N_{i+1}=\lg(2^{\lambda_i}/\beta_i)\leq \lambda_i$. This implies
\begin{align*}
    m_i&\leq
    4(1+\epsilon_{0,i})\frac{\lg p_i}{\lg p_{i+1}}\cdot
    \left(3+\frac{\lambda_i}{\lambda_{i+1}+1}\right)\cdot m_{i+1}\frac
{\lg p_{i+1}\lg\lg p_{i+1}}{\lg p_{i}\lg\lg p_{i}}
    +O(1)
\\
    &\leq4(1+\epsilon_{0,i})\left(3+\frac{\lambda_i}{\lambda_{i+1}+1}\right)
    \frac{\lg\lg  p_{i+1}}{\lg\lg p_i}m_{i+1}+O(1)\\
    \intertext{By
    Lemma~\ref{lemma:bound-lambda} we have
    $\left(3+\frac{\lambda_i}{\lambda_{i+1}+1}\right)\leq\frac{\lambda_i+9\lg\lambda_i}{\lambda_{i+1}+1}\leq\frac{\lambda_i+9\lg\lambda_i}{\lambda_{i+1}}$.
    Furthermore by statement
    ({\it ii\/}) from Lemma~\ref{lemma:bound-logprime} for $i>0$,
    we have $\lg\lg
    p_i\geq\lambda_i+\lg\lambda_i$,
    so that}
    m_i&\leq4\cdot (1+\epsilon_{0,i})
    \cdot
    \frac{\lambda_i+9\lg\lambda_i}{\lambda_i+\lg\lambda_i}
    \cdot
    \frac{\lg\lg p_{i+1}}{\lambda_{i+1}}
    \cdot
    m_{i+1}
    +O(1)
    \\
    &\leq4
    \cdot (1+\epsilon_{0,i})
    \cdot (1+\epsilon_{1,i})
    \cdot (1+\epsilon_{2,i})
    \cdot m_{i+1}
    +O(1).
\end{align*}
where we used again
    Lemma~\ref{lemma:bound-logprime} to bound $\lg\lg p_{i+1}$. This
    proves our
    claim.
\end{proof}

It is easy to convince oneself that the three quantities
$\epsilon_{0,i}$, $\epsilon_{1,i}$, and $\epsilon_{2,i}$ all tend to zero
as $\lambda_i$ grows (that is, as we deal with larger and larger input
numbers). The final asymptotic formula needs the following stronger
result, however.
\begin{lemma}
    \label{lemma:infprod}
    Let $\lambda_0$ be an arbitrarily large integer. Let $K$ be the first
    integer such that $\lambda_K<4$. 
    We have $K=\log^*\lambda_0+O(1)$. Furthermore,
    for $j=0,1,2$:
    $$\prod_{i=0}^{K-1}(1+\epsilon_{j,i})<\infty\rlap{\quad
    (independently of $K$)}$$
\end{lemma}
\begin{proof}
    The expression of $K$ follows from the inequality
    $\lambda'<3\log\lambda-1$ proved in Lemma~\ref{lemma:bound-lambda}. To
    see that, let $\Phi(\lambda)=3\lg\lambda-1$, defined for
    $\lambda\geq4$. Let $\Phi^*(x)$ be the function defined similarly to
    $\log^*$, by $\Phi^*(x)=0$ for $x<4$, and
    $\Phi^*(x)=1+\Phi^*(\Phi(x))$ otherwise. It is clear that
    $K\leq\Phi^*(\lambda_0)$. Now using the terminology defined in
    \cite[§5]{HavdHLe16}, the function $\Phi^*$ is an iterator for the
    logarithmically slow function $\Phi$. As such, it satisfies
    $\Phi^*(x)=\log^*x + O(1)$, which corresponds to our claim.
    
    To bound the product, it suffices to bound $\sum_i|\epsilon_{j,i}|$.
    Let $f_0(x)=\frac{4\lg x}{x}$,
    $f_1(x)=\frac{8\lg x}{x}$, and
    $f_2(x)=\frac{2+\lg x}{x}$, so that
    $\epsilon_{0,i}=f_0(\lambda_{i+1})$, 
    $\epsilon_{1,i}=f_1(\lambda_{i})$,
    $\epsilon_{2,i}=f_2(\lambda_{i+1})$.
    The functions $f_j$ are decreasing for $x\geq\exp(1)$. In particular,
    we have $\epsilon_{1,i}\leq f_1(\lambda_{i+1})$.
    Consider the sequence of real numbers
    defined by $u_0=3$, $u_1=4$, $u_2=6$, $u_3=27$, and $u_{k+1}=2^{u_k/3}$ for
    $k\geq3$. This sequences diverges to infinity. Independently of the starting
    value $\lambda_0$, we have
    \begin{align*}
        \lambda_{K}&\geq 3=u_0,\\
        \lambda_{K-1}&\geq 4=u_1,\\
        \lambda_{K-2}&\geq 6=u_2\rlap{ by observing
        Table~\ref{tab:lambda-lambdaprime},}\\
        \lambda_{K-3}&\geq 27=u_3\rlap{ again by 
        Table~\ref{tab:lambda-lambdaprime},}\\
        \lambda_{K-4}&\geq 2^{\lambda_{K-3}/3} \geq u_4\text{\ \ by
        Lemma~\ref{lemma:bound-lambda},}\\
        \lambda_{K-k} &\geq u_k\text{\ \ for all $k\leq K$.}
    \end{align*}
    This yields 
    \begin{align*}
        \sum_{i=0}^{K-1}
        \sum_{j=0}^2
        |\epsilon_{j,i}|
        &=
        \sum_{k=0}^{K-1}
        \sum_{j=0}^2
        |\epsilon_{j,K-1-k}|
        \leq
        \sum_{k=0}^{K-1}
        (f_0(\lambda_{K-k}) + f_1(\lambda_{K-1-k}) + f_2(\lambda_{K-k}))
        \\&
        \leq
        \sum_{k=0}^{K-1}
        \sum_{j=0}^2
        f_j(\lambda_{K-k})
        \leq
        \sum_{k=0}^{K-1}
        \sum_{j=0}^2
        f_j(u_k)
        \leq
        \sum_{k=0}^\infty
        \sum_{j=0}^2
        f_j(u_k).
    \end{align*}
    The latter sum converges to an absolute constant.
\end{proof}

\subsection{Complexity of integer multiplication}
\begin{theorem}
    \label{th:cplx}
    The complexity $\sM_{\text{new}}(n)$ of the algorithm presented in
    §\ref{sec:mulZ} to multiply $n$-bit integers is
    $$\sM_{\text{new}}(n)=O(n\cdot\log n\cdot 4^{\log^* n}).$$
\end{theorem}
\begin{proof}
    This theorem is a consequence of the results obtained thus far.
    Recall that
    in \algorithmref{algo:multfurer}, we have $N=O({n}/{(\lg n)^2})$
    and
    $2^\lambda=O(\lg n)$. The input size of
    \algorithmref{algo:mulZ} is $n_0=\Theta((\lg n)^2)$ bits. We have
    \begin{align*}
        O(N\log N\cdot2^\lambda\log n) &= O(({n}/{(\lg n)^2})(\lg
        n)^3) = O(n\log n).\\
        N(3\lceil \log_{{2^{\lambda+1}}} N \rceil +1)\cdot
        \sM_{\cR_0}
        &\leq O\left(\frac{n}{(\lg n)^2}\right)\cdot O\left(\frac{\lg
        n}{\lg\lg n}\right)\sM_{\cR_0}\\
        &\leq O(n\log n)\cdot \frac{\sM_{\cR_0}}{n_0
        (\lg n_0)}.
        \end{align*}
        {We thus have, using $\log p_0 = \Theta(n_0)$:}
        \begin{align*}
        \frac{\sM_{\text{new}}(n)}{n\log n} &= O(1) +
            O\left(\frac{\sM_{\cR_0}}{\lg p_0\lg\lg p_0
                }\right)
                        = O(1) + m_0
    \end{align*}
    using the notation of
    Proposition~\ref{prop:cplx-mi}.
    Let now $A$ be a constant bounding the $O(1)$ in
    Proposition~\ref{prop:cplx-mi}, let $C=A/3$, and let
    $e(i)=\prod_{0\leq j\leq 2}(1+\epsilon_{j,i})$. We have
    $4e(i)-1\geq 3$ so that $A\leq (4e(i)-1)C$.
    Proposition~\ref{prop:cplx-mi} implies 
    \begin{align*}
        m_i&\leq 4e(i) m_{i+1} +
    (4e(i)-1) C\\
        (m_i+C)&\leq4e(i) (m_{i+1}+C),
    \end{align*}
    so that we get $m_0=O(4^{\log^* n})$ by Lemma~\ref{lemma:infprod}.
    Finally, this gives 
    $$\frac{\sM_{\text{new}}(n)}{n\log n} = O(4^{\log^* n}).$$
\end{proof}

%
\section{Practical considerations}\label{sec:practical}

While our algorithm is mostly of theoretical interest, several points are
worth mentioning, as an answer to the natural question of its
practicality. Despite the title of this section, we are not reporting
data on an actual implementation of our algorithm, but rather
measurements that shed some light on its practical value.


\subsection{Adaptation of the asymptotically fast algorithm to practical sizes}
\label{sec:algo-adapt}

At the beginning of~§\ref{sec:mulZ}, we briefly alluded to a way to
multiply two $n$-bit integers: pick a generalized Fermat number (not a
priori prime) of the form $p_0=P(2^{\lambda_0},{\lambda_0})$, for
$\lambda_0$ such that $\lambda_0 2^{\lambda_0}\geq 2n$. Then use
\algorithmref{algo:mulR}. This does not work asymptotically because
computing roots of unity modulo $p_1=\smallerprime(p_0)$ cannot be done
deterministically with good complexity.  However, in practice, for say
$n\leq 2^{64}$, Table~\ref{tab:lambda-lambdaprime} and
Lemma~\ref{lemma:bound-logprime} imply that $p_1$ would then be at most a
2048-bit prime, for which both the primality proof and the computation of
roots can reasonably be assumed to be done once and for all. Therefore,
the stumbling blocks that are relevant for the asymptotic analysis need
not be considered as such for a practical implementation. This implies in
particular that resorting to \algorithmref{algo:multfurer}, as we do in
\algorithmref{algo:mulZ} for asymptotic reasons, is not needed in
practice.

Going further in this direction, we may in fact consider as a practical
instance of our algorithm the more general procedure that follows
Algorithm~\ref{algo:integer-mul:clump+eval} with $\cR_1=\ZnZ{p_1}$ as a
base ring, where $p_1$ is a generalized Fermat prime. The aforementioned
strategy can be regarded as Algorithm~\ref{algo:integer-mul:clump+eval}
with $\eta=2^{\lambda_0}$, $N=2^{\lambda_0}$ (still with $\lambda_0
2^{\lambda_0}\geq 2n$), at least in the case where
$\beta=\batchsize(p_0)=1$.

Another alteration that we wish to make in practice is that our top-level
multiplication need not use a negacyclic transform: whether we compute a
product modulo $2^{2n}+1$ or $2^{2n}-1$ makes no difference when both
inputs are less than or equal to $2^n-1$. On the other hand, a ``full''
DFT of length $N$ instead of a Half-DFT saves $3N$ multiplications in
the base ring, which is not entirely negligible.

Finally, we note that for all sizes of practical interest, arithmetic in
$\cR_1=\ZnZ{p_1}$ will not be done with a Fourier-transform-based
algorithm, because $p_1$ is only of very moderate size.

Taking into account all the remarks above, the only link that remains between the practical
procedure that we envision and the algorithms (in particular,
\algorithmref{algo:mulR}) described in this article is
that $p_1$ is a generalized Fermat prime. The developments in this
article show that computing with generalized Fermat prime is
asymptotically feasible, and yields a good complexity.

\subsection{Parameter choices for various input sizes}
In this section, we consider various input sizes $n$, and various
candidate generalized Fermat primes $p_1=\gfprime{r_1}{\lambda_1}$. For
combinations of these, we find values $\eta$ and $N$ (both powers of two)
such that Algorithm~\ref{algo:integer-mul:clump+eval} works. Let us
briefly recall its structure: we write both $n$-bit integer inputs $a$
and $b$ in radix $\eta$, or equivalently as the evaluations at $\eta$ of
two polynomials of degree less than $N/2$. We multiply these polynomials
in $\cR_1[x]$. For this, we compute full $N$-point DFTs, then a pointwise
product, and finally an inverse DFT. Arithmetic in $\cR_1$, as in
§\ref{sec:algo}, uses representation in radix $r_1$. For this procedure
to correctly compute the integer product $a\cdot b$, the following
conditions must hold:
$$\left\{
    \begin{array}{ll}
        N\eta^2\leq p_1 & \text{(no overflow occurs in $\cR_1$)},\\
        2^{2n} \leq \eta^N-1 & \text{(correct computation of the product of two $n$-bit
        integers)},\\
        N \mid p_1-1 &\text{(a principal $N$-th root of unity exists in $\cR_1$)}.
    \end{array}\right.
    $$
In particular, $N$ is the smallest power of two above $2n/\lg \eta$.
When choosing $\eta$ and $N$ subject to the conditions above, we have
some freedom. Ultimately, we wish to minimize the number of
multiplications in $\cR_1$, because we expect those to form the largest
part of the computation time. More precisely, we wish to minimize the
overall cost $(3E(N)+N)\sM_{\cR_1}$ of \emph{expensive} multiplications as
introduced in §\ref{sec:furerbounds} ($\sM_{\cR_1}$ denotes the cost of
one expensive multiplication in $\cR_1$; we add $N$ because of the
pointwise products, and not $4N$ since here we do not use a half-DFT).
Using the expression of $E(N)$, a rough estimate of the quantity to
minimize is $\frac{n}{\lg\eta}\frac{\lg n}{\lambda_1 + 1}\sM_{\cR_1}$,
therefore for $n$ constant we try to minimize
$$Q\approx\frac{\sM_{\cR_1}}{\lambda_1\lg\eta}.$$

Thus, there is a trade-off to determine: when $\eta$ grows, larger primes
$p_1$ have to be used: ${\sM_{\cR_1}}$ increases, while $\frac{1}{\lg
\eta}$ decreases.  Since the cost $\sM_{\cR_1}$ is given by the bit
length of the prime $p_1$, the $\eta$ that we choose should be the
largest $\eta$ for which $p_1$ is valid (as per the first of the three
conditions above). The number of expensive multiplications for various
input sizes $n$ and primes $p_1$ is reported in
Table~\ref{table:tradeoff}.
We added in Table~\ref{table:tradeoff} the
additional constraint that $\lg\eta$ be a multiple of the machine word
size, to the extent possible (since $N$ must be a power of two anyway,
this constraint has no impact).

\begin{table}
	\begin{center}
            \begin{adjustbox}{max width=\textwidth}
    \input{table_tradeoff.magmadata.txt}

    \end{adjustbox}
\end{center}
\caption{Estimated lower bound for the total cost of expensive
    multiplications in our algorithm depending on the prime used. Timings
    are based on the multiplication count and the measured time for the
    Kronecker-Schönhage bit length in the fifth column, on an Intel Xeon
    E7-4850v3 CPU (2.20GHz).}
\label{table:tradeoff}
\end{table}

\subsection{Cost of multiplications in the underlying ring}
We now turn to the two last columns of Table~\ref{table:tradeoff}. Our
goal is to obtain a coarse lower bound on the time we expect our
algorithm to take.
Arithmetic in $\cR_1$, and in particular multiplication, is our main
focus. Elements of $\cR_1$ are represented in radix $r_1$. We avoid the
conversion between radix $r_1$ and binary representation by using
Kronecker substitution: an element of $\cR_1$, represented as a
$2^{\lambda_1}$-uple of integers in $\closedopeninterval{0,r_1}$, is
transformed into an integer of bit length $$k=(2\lceil\lg
r_1\rceil+\lambda_1)\cdot 2^{\lambda_1}.$$ Multiplication in $\cR_1$ is
then done by multiplying these integers modulo $2^k+1$ (we deal with
signs in the same way as in \algorithmref{algo:mulR}). We ignore
the cost of converting this product back to radix $r_1$. This is
likely to be at the very least a significant source of inaccuracy in our
lower bounds.

The fifth column of Table~\ref{table:tradeoff} reports the bit length $k$
introduced above, for the various generalized Fermat primes chosen. Based
on this bit length, we determined experimentally on a target machine
(Intel Xeon E7-4850v3 CPU clocked at 2.20GHz) the time taken by the
function \verb|mpn_mul|
in the GMP library~\cite{GMP16}, thereby giving a
lower bound on the multiplication time in $\cR_1$.  We multiplied this
lower bound by the number of expensive multiplication reported on the
fourth column of Table~\ref{table:tradeoff}, from which we deduced a
lower bound on the multiplication time for $n$-bit integers using our
algorithm.

The determination of the bit length above led us to restrict the set of
generalized Fermat primes to consider: two such primes
that lead to identical bit length lead to an identical time for
internal multiplications. Therefore, we favor the largest generalized
Fermat prime for each value of the Kronecker-Schönhage bit length $k$
above. In our choice, we also favored primes such that $r_1$ has largest
$2$-valuation among the candidate values (e.g.\ both $984^{16}+1$ and
$1018^{16}+1$ are primes, but we experimented with the former because the
latter only allows a maximum transform length of $2^{16}$).

We deduce from Table~\ref{table:tradeoff} that for realistic sizes,
choosing the prime $p_1$ appropriately can lead to a speed-up of the
order of~$2$ to~$4$, with all the necessary words of caution: as
mentioned above, we deliberately omitted some conversion costs that are
unlikely to be negligible in practice, and also our measurements are done
with all operands in cache memory, which is quite probably optimistic.

\subsection{Comparison with Schönhage-Strassen}

Let us compare approximatively the cost of Schönhage-Strassen's algorithm
to our algorithm. We can do two things. At least up to some size, we can
run GMP's implementation of the Schönhage-Strassen algorithm,
and obtain an actual computation time. Or we can do as we did in
Table~\ref{table:tradeoff}: count the number of small multiplications
involved, and measure their cost. We did both, because the latter
approach, which inherently gives a lower bound, is a fairer comparison
given that a lower bound is all that we have in
Table~\ref{table:tradeoff}.

Roughly speaking, a Schönhage-Strassen multiplication of two $2^n$-bit
integers involves $2^{\lfloor (n+1)/2\rfloor}$ multiplications of
$2^{1+\lceil (n+1)/2\rceil}$-bit modular integers.  In truth, a
well-tuned implementation of the Schönhage-Strassen algorithm uses all
sorts of optimizations that are well outside the scope of this article
(see e.g.\ \cite{Gaudry:2007:GIS:1277548.1277572}), so that this is a
crude estimate.

In Table~\ref{table:comparison-ss}, we report how our lower bound
compares to the lower bound that we obtain in this way on the running
time of the Schönhage-Strassen algorithm. As we did in
Table~\ref{table:tradeoff}, the fourth column is computed by determining
experimentally the individual cost of each of the underlying
multiplications. For this, we timed GMP's internal routine
\verb|mpn_mul_fft|, as it is called in the implementation. The fifth
column of Table~\ref{table:comparison-ss} indicates the real computation
time, measured experimentally (we modified GMP's internal
\verb|mp_size_t| type to go beyond 31 bits). Our measurements were limited
by core memory, since the product of two $2^{40}$-bit integers took 1.3TB
of RAM. The comparison with the previous column shows that our lower
bound on the Schönhage-Strassen time is within a factor of two of the real computation time, which is
acceptable.

\begin{table}
    \begin{center}
        \begin{adjustbox}{max width=\textwidth}
            \input{table_comparison_ss.magmadata.txt}

        \end{adjustbox}
        \caption{
            \label{table:comparison-ss}
        Comparison of lower bounds on the running time of the
        Schönhage-Strassen algorithm and the practical algorithm
        described in~\ref{sec:algo-adapt}. The right half is from
        Table~\ref{table:tradeoff}. Timings measured on an Intel Xeon
        E7-4850v3 CPU (2.20GHz).}
    \end{center}
\end{table}

We conclude from Table~\ref{table:comparison-ss} that an
implementation of our algorithm will unlikely beat an implementation of
the Schönhage-Strassen algorithm for sizes below $2^{40}$.
Above $2^{40}$, the ratio of our lower bounds is only slightly more than
two.
We may
speculate that an optimized
implementation could compensate this gap.


One of the arguments in favor of our algorithm is that its memory
locality is likely much better, because of the shallow recursion.

A direction to consider for optimization can be to improve on the time needed for
internal multiplications. For example, we may represent elements of $\cR_1$ in radix
$r_1^2$ instead of $r_1$. In some cases, it might lead to a smaller bit
length, at the expense of some extra conversion costs. For example for
$p_1=1984^{16}+1$, working in radix $1984^2$ leads to polynomials
of length $8$, and a bit length of $8\cdot48=384$ bits, instead of $416$
bits (see Table~\ref{table:tradeoff}).
Another possibility is to use
the multipoint Kronecker substitution proposed by
Harvey in~\cite{DBLP:journals/corr/abs-0712-4046}.  For this same example,
evaluating at
$+2^{24}$ and $-2^{24}$, we can compute the product via two
multiplications
of two 192-bit integers, which might be faster. For both ideas however,
we have not 
taken into account the
conversion costs, and it seems difficult to be very confident about the
induced benefit.

\section{Conclusions}\label{sec:conclusion}
Our algorithm follows Fürer's perspective, and improves on the cost of
the multiplications in the underlying ring. Although of similar
asymptotic efficiency, it therefore differs from the algorithm
in~\cite{HavdHLe16}, which is based on Bluestein's chirp transform,
Crandall-Fagin reduction, computations modulo a Mersenne prime, and
balances the costs of the ``expensive'' and ``cheap'' multiplications.

It is interesting to note that both algorithms rely on hypotheses related
to the repartition of two different kinds of primes.  It is not clear
which version is the most practical, but our algorithm avoids the use of
bivariate polynomials and seems easier to plug in a classical
radix-${2^{\lambda}}$ FFT by modifying the arithmetic involved. The only
additional cost we have to deal with is the question of the decomposition
in radix~$r$, and the computation of the modulo, which can be improved
using particular primes. However, we do not expect it to beat
Schönhage-Strassen for integers of size below $2^{40}$ bits.

A natural question arises: can we do better? The factor $4^{\log^*  n}$
comes from the direct and the inverse FFT we have to compute at each
level of recursion, the fact that we have to use some zero-padding each
time, and of course the recursion depth, which is $\log^*  n + O(1)$.

Following the same approach, it seems hard to improve on any of the
previous points. Indeed, the evaluation-interpolation paradigm suggests a
direct and an inverse FFT, and getting a recursion depth of
$\frac{1}{2}\log^* n + O(1)$ would require a reduction from $n$ to
$\log\log n$ at each step.


\section*{Acknowledgements}

The authors are indebted to the anonymous referee, whose careful reading
greatly helped enhance the presentation of this article.

\bibliographystyle{amsalpha}
\bibliography{furer}

\appendix
%
\section{Proof of Proposition~\ref{prop:lowerbound-C}}
\label{appendix:proofconjc}
\conjC*

\newcommand*\dif{\mathop{}\!\mathrm{d}}

\begin{proof}
    We prove that there exists an absolute constant $C>0$ such that
    $C_\lambda\geq \frac{C}{\lambda}$ for any $\lambda\geq 1$, where
    $C_\lambda$ is defined as in Lemma~\ref{lemma:exists-gfp}.

    The idea is to rely on the proof of the main theorem of~\cite[§2]{pomerance77}, and to use the main result of~\cite{elliott2007}
    for arithmetic progressions with ``powerful moduli'', since we consider arithmetic progressions $(q\cdot k + r)_k$
    where $q$ is a power of two.

    Let $\mathcal{P}(x)$ be the set of primes smaller than $x$, and
    extend the notation of
    Lemma~\ref{lemma:exists-gfp}  to define
$$C_\lambda(x)=\frac12\prod_{{p\in\cP(x)
}}\frac{1-{\chi_\lambda(p)}/p}{1-1/p}.$$

    Throughout this appendix, we use the shorthand notation
    $q=2^{\lambda+1}$.
 Let $\pi(x,q,r)$
 be the number of primes $\leq x$ congruent to $r$ mod $q$.
    Let 
    $G(x)=\pi(x,q,1)$.
    We will use twice the Brun-Titchmarsh inequality, which says that
    $$G(x)=\pi(x,q,1)\leq2(x/\phi(q))/\log(x/q)=4x/(q\log(x/q)).$$
    Let now
    $g(t)=G(t)-G(t-q)$.
    By construction, $G$ and $g$ are constant
    on intervals $\closedopeninterval{1+qi,1+qi+q}$ for any integer $i\geq0$,
    and $g(t)$ is equal to
    to
    $1$ or $0$ on that interval depending on whether
    $1+qi$ is prime or not. Hence
    $$g(1+qi)=\frac1q\int_{1+qi}^{1+qi+q}g(t)\dif t\quad\text{and}\quad
    G(1+qi)=\frac1q\int_0^{1+qi+q}g(t)\dif t.$$

    Let $F(x)=-\log(1-2^\lambda/x)=-\log(1-q/(2x))$, which is a
    decreasing, convex, and nonnegative function defined for $x>q/2$.
    Furthermore, since $\lambda\geq0$, for $x\geq 1+q/2$ we have
    $F(x)\leq\log 2$.  Our goal is to find an asymptotic lower bound for
    the (logarithm of the) numerator of $C_\lambda(x)$ for large $x$ (we
    impose $x\geq 1+2q$ below).
    Equivalently, we seek an upper bound for $\cS(x)=\sum_{i=1}^{N}
        F(1+qi)g(1+qi)
$, where
    we set
    $N=\left\lfloor(x-1)/q\right\rfloor$.
    Throughout the proof below,
    implicit constants 
    $O(1)$
    are uniform on $\lambda$ ---possibly for $x$ larger than some bound
    that depends on $\lambda$, but that is not an issue since
    $C_\lambda=\lim_{x\rightarrow\infty}C_\lambda(x)$.
    \begin{align*}
        \cS(x) &
        = \sum_{i=1}^{N} F(1+qi)g(1+qi)
        = \sum_{i=1}^{N} {F(1+qi)\frac1q\int_{1+qi}^{1+qi+q}g(t)\dif t}
        \\
        &
        \leq\frac1q\sum_{i=1}^{N}\int_{1+qi}^{1+qi+q}F(t-q/2)g(t)\dif t
        \rlap{\quad (because $F$ is convex)}
        \\&
        \leq\frac1q\int_{1+q}^{x+q}F(t-q/2)g(t)\dif t +
        \underbrace{\frac1q\int_{x+q}^{1+qN+q}
        F(t-q/2)g(t)\dif t }_{O(1)}
        \\
        &\leq
        {F(x+q/2)G(x)}-F(1+q/2)G(1+q)-\int_{1}^{x}
        F'(t+q/2)G(t)\dif t +
        O(1)
        .
        \intertext{%
        Since $F(x+q/2)\leq\frac q{2x}$ and
        $G(x)\leq 4x/(q\log x/q))$, the first
        summand is bounded by $2/\log 2$ for $x\geq 2q$.
        Since $G(1+q)\leq1$ and $G(t)=0$ for $t<2$ we have:}
        \cS(x)&
        \leq O(1) + \int_{2}^{x}\frac{q\pi(t,q,1)}{t(2t+q)}\dif t
        \leq O(1) + \int_{2}^{x}\frac{q\pi(t,q,1)}{2t^2}\dif t.
    \end{align*}

    Elliott~\cite{elliott2007} proved
    a theorem that relates $\pi(x,q,r)$ to its asymptotic estimate
    We state a very weak form of it,
    namely
%
%
    that
    there exists an absolute constant $K$ such that for any $\lambda\geq 0$ and $t$ such that 
    $$\min(t^{1/3}\exp(-(\log\relax\log t)^3),t^{1/2}\exp(-8\log\relax\log t)) \geq q,$$
    we have
    \begin{equation}
    \label{eq:ineqelliott}
        \left|\pi(t,q,1)-\frac{2 t}{q\log t}\right| < \frac{K t}{q(\log t)^2}.
    \end{equation}

    The condition above on $t$ can be
    simplified.
    There exists an absolute constant $H$ such that for any $x>1$
    $$\min(x^{1/3}\exp(-(\log\relax\log x)^3),x^{1/2}\exp(-8\log\relax\log x)) \geq (H x)^{1/4}.$$
    Thus, for $t \geq q^4/H$, 
    Equation~\eqref{eq:ineqelliott} holds.
    We rewrite the
    upper bound on $\cS(x)$:
    \begin{align*}
        \cS(x)&\leq O(1)
        +\underbrace{\int_2^{q^4/H}\frac{q\pi(t,q,1)}{2t^2}\dif
        t}_{I_0(\lambda)} 
        +\underbrace{\int_{q^4/H}^x\frac{q\pi(t,q,1)}{2t^2}\dif
        t}_{I_1(\lambda,x)}.
    \end{align*}
    For $I_0(\lambda)$, by Brun-Titchmarsh we have
    \begin{align*}
        I_0(\lambda)&\leq\int_2^{q^4/H}\frac{2}{t\log(t/q)}\dif t
        \leq\int_2^{q^3/H}\frac{2}{u\log(u)}\dif u
        \\&
        \leq  2\log\relax\log(q^3/H)\leq 2\log\lambda+O(1).\\
    \end{align*}
    We use Elliott's theorem to bound $I_1(\lambda,x)$ (using the notations of
    Equation~\eqref{eq:ineqelliott}):
    \begin{align*}
    \left|I_1(\lambda,x)-\int_{q^4/H}^{x} \frac{1}{t\log t}\mathrm{d}t\right|
        &\leq \int_{q^4/H}^{x}{\frac{K}{t(\log t)^2}\mathrm{d}t}\\
        &\leq -\frac K{\log x} + \frac K{\log(q^4/H)}
        =O(1)\\
        \llap{so that }I_1(\lambda,x)&\leq\log\relax\log x-\log\relax\log(q^4/H)+O(1)\\
        &\leq\log\relax\log x-\log\lambda+O(1).
    \end{align*}
    Combining the bounds on $I_0$ and $I_1$, we have obtained:
    $$\cS(x)\leq\log\relax\log x + \log\lambda + O(1).$$
    \medskip

    The lower bound on $C_\lambda(x)$ follows: indeed, we have
    \begin{align*}
        -\log C_\lambda(x) &= {\sum_{p \in
        \mathcal{P}(x)}{\log\left(1-\frac{1}{p}\right)}} + \cS(x),\\
        &\leq \left(-\gamma -\log\relax\log x + o(1)\right) + \left(\log\relax\log x + \log\lambda +
        O(1)\right),\\
        \log C_\lambda(x) &\geq O(1) - \log\lambda.
    \end{align*}
        Hence $C_\lambda(x) \geq A/\lambda$ for some absolute constant
        $A$, and $x$ large enough. It follows that $C_\lambda\geq
        A/\lambda$, as claimed.
    We notice that the multiplier affecting $\log\lambda$ above, and
    hence the exponent of $\lambda$ in our lower bound, can be
    directly traced to the use of the Brun-Titchmarsh inequality in
    bounding $I_0(\lambda)$.
\end{proof}
\end{document}

%% file: furer-macros.tex
\usepackage[T1]{fontenc}
\usepackage[utf8]{inputenc}
\usepackage{algorithm}
\usepackage{hyperref}
\usepackage{tikz}
\usepackage{algorithmicx}
\usepackage{algpseudocode}
\usepackage{adjustbox}
\usepackage{thmtools,thm-restate}

\newtheorem{theorem}{Theorem}[section]
\newtheorem*{theorem*}{Theorem}
\newtheorem{lemma}[theorem]{Lemma}
\newtheorem{proposition}[theorem]{Proposition}
\newtheorem{hypothesis}[theorem]{\textbf{Hypothesis}}
\newtheorem*{recallhypothesis}{\textbf{Hypothesis~\ref{conj:second}}}
\newtheorem{notation}[theorem]{\textbf{Notation}}

\theoremstyle{definition}
\newtheorem{definition}[theorem]{Definition}

\theoremstyle{remark}

\numberwithin{equation}{section}
\makeatletter
\let\olog\log

\def\olg{\log_2}
\def\mlog#1{\olog^{(#1)}}
\def\mlg#1{\olg^{(#1)}}

\newcount\@logcount
\def\@zlg{\ifnum\@logcount=1\olg\else
\edef\reserved@a{\noexpand\mlg{\the\@logcount}}\reserved@a
\fi}
\def\@lginner{\ifx\reserved@a\lg\def\next\lg{\@lg}\else\let\next\@zlg\fi\next}
\def\@lg{\advance\@logcount1\futurelet\reserved@a\@lginner}
\def\@zlog{\ifnum\@logcount=1\olog\else
    \edef\reserved@a{\noexpand\mlog{\the\@logcount}}\reserved@a
\fi}
\def\@loginner{\ifx\reserved@a\log\def\next\log{\@log}\else\let\next\@zlog\fi\next}
\def\@log{\advance\@logcount1\futurelet\reserved@a\@loginner}
\DeclareRobustCommand{\lg}{\@logcount0\@lg}
\DeclareRobustCommand{\log}{\@logcount0\@log}
\makeatother
\makeatletter
\RequirePackage{marginnote}

\makeatother
\makeatletter
\def\@doletters#1{\def\@b{#1}\ifx\@b\@c\let\@a\egroup\else\@d{#1}\fi\@a}
\def\doletters#1#2{\bgroup\def\@d##1{\global\@namedef{#1##1}{\ensuremath{#2{##1}}}}\def\@a{\@doletters}\def\@c{;}\@a}
\doletters c\mathcal  ABCDEFGHIJKLMNOPQRSTUVWXYZ;
\doletters s\mathsf   ABCDEFGHIJKLMNOPQRSTUVWXYZ
                    {BW}{BL}{LD}{DL}{RSA}{NFS}{SNFS}{GNFS};
\doletters b\mathbb   ABCDEFGHIJKLMNOPQRSTUVWXYZ;
\doletters {fk}\mathfrak ABCDEFGHIJKLMNOPQRSTUVWXYZ
                         abcdefghijklmnopqrstuvwxyz;
\doletters {bf}\mathbf   ABCDEFGHIJKLMNOPQRSTUVWXYZ
                         abcdefghijklmnopqrstuvwxyz;
\makeatother
\newcommand{\rquo}[2]{\leavevmode\ensuremath{\kern-.1em\raise.2ex\hbox{$#1$}\kern-.1em/\kern-.1em\lower.25ex\hbox{$#2$}}}
\let\quotient\rquo

\algnewcommand\Input{\State\textbf{Input:}\space}
\algnewcommand\Output{\State\textbf{Output:}\space}
\algnewcommand\Remark{\State\textbf{Remark:}\space}

\algrenewcommand\textproc{\textsf}
\def\closedintervalZ#1{\mathopen{[\![}#1\mathclose{]\!]}}
\def\closedopenintervalZ#1{\mathopen{[\![}#1\mathclose{[\![}}
\def\closedopeninterval#1{\mathopen{[}#1\mathclose{)}}
\DeclareMathOperator{\smallerprime}{smallerprime}
\DeclareMathOperator{\batchsize}{batchsize}
\DeclareMathOperator{\smallerring}{smallerring}

\DeclareMathOperator{\li}{li}
\DeclareMathOperator{\DFT}{DFT}
\DeclareMathOperator{\HDFT}{Half-DFT}
\def\gfprime#1#2{\ensuremath{{#1}^{2^{#2}}+1}}
\makeatletter
\def\@doletters#1{\def\@b{#1}\ifx\@b\@c\let\@a\egroup\else\@d{#1}\fi\@a}
\def\doletters#1#2{\bgroup\def\@d##1{\global\@namedef{#1##1}{\ensuremath{#2{##1}}}}\def\@a{\@doletters}\def\@c{;}\@a}
\doletters c\mathcal  ABCDEFGHIJKLMNOPQRSTUVWXYZ;
\doletters s\mathsf   ABCDEFGHIJKLMNOPQRSTUVWXYZ
                    {BW}{BL}{LD}{DL}{RSA}{NFS}{SNFS}{GNFS};
\doletters b\mathbb   ABCDEFGHIJKLMNOPQRSTUVWXYZ;
\doletters {fk}\mathfrak ABCDEFGHIJKLMNOPQRSTUVWXYZ
                         abcdefghijklmnopqrstuvwxyz;
\doletters {bf}\mathbf   ABCDEFGHIJKLMNOPQRSTUVWXYZ
                         abcdefghijklmnopqrstuvwxyz;

\def\ZnZ#1{\ensuremath{\rquo{\bZ}{#1\bZ}}}

\def\newalgname#1#2{\global\@namedef{#1%
}{#2}}
\DeclareRobustCommand{\algorithmlabel}[1]{%
    \label{#1}%
    \@bsphack
        \protected@write\@auxout{}{\string\newalgname{algname@#1}{\currentalgorithm}}%
    \@esphack
}%
\def\algorithmref#1{Algorithm~\ref{#1}
(\textproc{\@nameuse{algname@#1}})}
\newenvironment{myalgorithm}[1]{%
    \def\currentalgorithm{#1}%
    \begin{algorithm}%
}{%
    \end{algorithm}%
}
\algblock{BeginBlock}{EndBlock}
\algnotext{BeginBlock}
\algnotext{EndBlock}

\makeatother

%% file: table_tradeoff.magmadata.txt
\begin{tabular}{|c|c|c|c|r|c|}
\hline\multicolumn{6}{|c|}{bit length of both operands: $2^{30}$} \\
\hline $p_1$ & $\eta$ & $N$ & $3E(N)+N$ & \multicolumn{1}{c|}{bit length of K.S.} & lower bound \\
\hline$\mathbf{984^{16}+1}$ & $\mathbf{2^{64}}$ & $\mathbf{2^{25}}$ & $\mathbf{2^{25}\cdot (16=3\cdot 5+1)}$ & $\mathbf{(2\cdot 10+4)\cdot 16=384}$ & $\mathbf{2.68\cdot 10^{1}\ \text{s}}$\\

\hline${1984^{16}+1}$ & ${2^{64}}$ & ${2^{25}}$ & ${2^{25}\cdot (16=3\cdot 5+1)}$ & ${(2\cdot 11+4)\cdot 16=416}$ & ${3.44\cdot 10^{1}\ \text{s}}$\\

\hline${4016^{16}+1}$ & ${2^{64}}$ & ${2^{25}}$ & ${2^{25}\cdot (16=3\cdot 5+1)}$ & ${(2\cdot 12+4)\cdot 16=448}$ & ${3.44\cdot 10^{1}\ \text{s}}$\\

\hline${448^{32}+1}$ & ${2^{128}}$ & ${2^{24}}$ & ${2^{24}\cdot (13=3\cdot 4+1)}$ & ${(2\cdot 9+5)\cdot 32=736}$ & ${3.82\cdot 10^{1}\ \text{s}}$\\

\hline${884^{32}+1}$ & ${2^{128}}$ & ${2^{24}}$ & ${2^{24}\cdot (13=3\cdot 4+1)}$ & ${(2\cdot 10+5)\cdot 32=800}$ & ${4.45\cdot 10^{1}\ \text{s}}$\\

\hline${412^{64}+1}$ & ${2^{256}}$ & ${2^{23}}$ & ${2^{23}\cdot (13=3\cdot 4+1)}$ & ${(2\cdot 9+6)\cdot 64=1536}$ & ${7.57\cdot 10^{1}\ \text{s}}$\\

\hline${506^{128}+1}$ & ${2^{512}}$ & ${2^{22}}$ & ${2^{22}\cdot (10=3\cdot 3+1)}$ & ${(2\cdot 9+7)\cdot 128=3200}$ & ${9.94\cdot 10^{1}\ \text{s}}$\\

\hline\multicolumn{6}{|c|}{bit length of both operands: $2^{40}$} \\
\hline $p_1$ & $\eta$ & $N$ & $3E(N)+N$ & \multicolumn{1}{c|}{bit length of K.S.} & lower bound \\
\hline${984^{16}+1}$ & ${2^{32}}$ & ${2^{36}}$ & ${2^{36}\cdot (25=3\cdot 8+1)}$ & ${(2\cdot 10+4)\cdot 16=384}$ & ${8.57\cdot 10^{4}\ \text{s}}$\\

\hline$\mathbf{1984^{16}+1}$ & $\mathbf{2^{64}}$ & $\mathbf{2^{35}}$ & $\mathbf{2^{35}\cdot (22=3\cdot 7+1)}$ & $\mathbf{(2\cdot 11+4)\cdot 16=416}$ & $\mathbf{4.84\cdot 10^{4}\ \text{s}}$\\

\hline${4016^{16}+1}$ & ${2^{64}}$ & ${2^{35}}$ & ${2^{35}\cdot (22=3\cdot 7+1)}$ & ${(2\cdot 12+4)\cdot 16=448}$ & ${4.84\cdot 10^{4}\ \text{s}}$\\

\hline${448^{32}+1}$ & ${2^{64}}$ & ${2^{35}}$ & ${2^{35}\cdot (19=3\cdot 6+1)}$ & ${(2\cdot 9+5)\cdot 32=736}$ & ${1.14\cdot 10^{5}\ \text{s}}$\\

\hline${884^{32}+1}$ & ${2^{128}}$ & ${2^{34}}$ & ${2^{34}\cdot (19=3\cdot 6+1)}$ & ${(2\cdot 10+5)\cdot 32=800}$ & ${6.66\cdot 10^{4}\ \text{s}}$\\

\hline${412^{64}+1}$ & ${2^{256}}$ & ${2^{33}}$ & ${2^{33}\cdot (16=3\cdot 5+1)}$ & ${(2\cdot 9+6)\cdot 64=1536}$ & ${9.54\cdot 10^{4}\ \text{s}}$\\

\hline${506^{128}+1}$ & ${2^{512}}$ & ${2^{32}}$ & ${2^{32}\cdot (13=3\cdot 4+1)}$ & ${(2\cdot 9+7)\cdot 128=3200}$ & ${1.32\cdot 10^{5}\ \text{s}}$\\

\hline\multicolumn{6}{|c|}{bit length of both operands: $2^{50}$} \\
\hline $p_1$ & $\eta$ & $N$ & $3E(N)+N$ & \multicolumn{1}{c|}{bit length of K.S.} & lower bound \\
\hline${984^{16}+1}$ & ${2^{32}}$ & ${2^{46}}$ & ${2^{46}\cdot (31=3\cdot 10+1)}$ & ${(2\cdot 10+4)\cdot 16=384}$ & ${1.09\cdot 10^{8}\ \text{s}}$\\

\hline$\mathbf{1984^{16}+1}$ & $\mathbf{2^{64}}$ & $\mathbf{2^{45}}$ & $\mathbf{2^{45}\cdot (28=3\cdot 9+1)}$ & $\mathbf{(2\cdot 11+4)\cdot 16=416}$ & $\mathbf{6.31\cdot 10^{7}\ \text{s}}$\\

\hline${4016^{16}+1}$ & ${2^{64}}$ & ${2^{45}}$ & ${2^{45}\cdot (28=3\cdot 9+1)}$ & ${(2\cdot 12+4)\cdot 16=448}$ & ${6.31\cdot 10^{7}\ \text{s}}$\\

\hline${448^{32}+1}$ & ${2^{64}}$ & ${2^{45}}$ & ${2^{45}\cdot (25=3\cdot 8+1)}$ & ${(2\cdot 9+5)\cdot 32=736}$ & ${1.54\cdot 10^{8}\ \text{s}}$\\

\hline${884^{32}+1}$ & ${2^{128}}$ & ${2^{44}}$ & ${2^{44}\cdot (25=3\cdot 8+1)}$ & ${(2\cdot 10+5)\cdot 32=800}$ & ${8.97\cdot 10^{7}\ \text{s}}$\\

\hline${412^{64}+1}$ & ${2^{256}}$ & ${2^{43}}$ & ${2^{43}\cdot (22=3\cdot 7+1)}$ & ${(2\cdot 9+6)\cdot 64=1536}$ & ${1.34\cdot 10^{8}\ \text{s}}$\\

\hline${506^{128}+1}$ & ${2^{512}}$ & ${2^{42}}$ & ${2^{42}\cdot (19=3\cdot 6+1)}$ & ${(2\cdot 9+7)\cdot 128=3200}$ & ${1.98\cdot 10^{8}\ \text{s}}$\\

\hline
\end{tabular}

%% file: table_comparison_ss.magmadata.txt
\begin{tabular}{|c|c|c|c|c|c|c|c|c|}
\hline
& \multicolumn{4}{c|}{Schönhage-Strassen algorithm}
& \multicolumn{4}{c|}{§\ref{sec:algo-adapt}}\\
\hline {bit length}& {\parbox{5em}{\centering \#internal\\products}}& {\parbox{5em}{\centering internal\\bit length}}& {lower bound}& {real time}
& {\parbox{5em}{\centering \#internal\\products}}& prime& {\parbox{5em}{\centering internal\\bit length}}& {lower bound}\\
\hline$2^{30}$ & $2^{15}$ & $\approx 2^{17}$ & $9.73\cdot 10^{0}\ \text{s}$ & $1.50\cdot 10^{1}\ \text{s}$ & $2^{25}\cdot 16$ & $984^{16}+1$ & $384$ & $2.68\cdot 10^{1}\ \text{s}$\\
\hline$2^{35}$ & $2^{18}$ & $\approx 2^{19}$ & $3.70\cdot 10^{2}\ \text{s}$ & $6.03\cdot 10^{2}\ \text{s}$ & $2^{30}\cdot 19$ & $984^{16}+1$ & $384$ & $1.02\cdot 10^{3}\ \text{s}$\\
\hline$2^{40}$ & $2^{20}$ & $\approx 2^{22}$ & $1.63\cdot 10^{4}\ \text{s}$ & $3.04\cdot 10^{4}\ \text{s}$ & $2^{35}\cdot 22$ & $1984^{16}+1$ & $416$ & $4.84\cdot 10^{4}\ \text{s}$\\
\hline$2^{45}$ & $2^{23}$ & $\approx 2^{24}$ & $7.90\cdot 10^{5}\ \text{s}$ & --- & $2^{40}\cdot 25$ & $1984^{16}+1$ & $416$ & $1.76\cdot 10^{6}\ \text{s}$\\
\hline$2^{50}$ & $2^{25}$ & $\approx 2^{27}$ & $2.88\cdot 10^{7}\ \text{s}$ & --- & $2^{45}\cdot 28$ & $1984^{16}+1$ & $416$ & $6.31\cdot 10^{7}\ \text{s}$\\
\hline$2^{55}$ & $2^{28}$ & $\approx 2^{29}$ & $1.05\cdot 10^{9}\ \text{s}$ & --- & $2^{50}\cdot 31$ & $4016^{16}+1$ & $448$ & $2.23\cdot 10^{9}\ \text{s}$\\
\hline$2^{60}$ & $2^{30}$ & $\approx 2^{32}$ & $3.44\cdot 10^{10}\ \text{s}$ & --- & $2^{55}\cdot 34$ & $4016^{16}+1$ & $448$ & $7.84\cdot 10^{10}\ \text{s}$\\
\hline
\end{tabular}